\documentclass[12pt,preprint]{aastex}

\def\meth {CH$_3$OH}
\def\hho  {H$_2$O}
\def\SgrA {Sgr~A*}

\def\p    {\phantom{0}}
\def\err  {\phantom{$\pm$0.0}}

\def\kms  {km~s$^{-1}$}
\def\kmsperkpc  {km~s$^{-1}$~kpc$^{-1}$}
\def\masy {mas~y$^{-1}$}
\def\uasy {$\mu$as~y$^{-1}$}
\def\uas  {$\mu$as}
\def\deg  {\ifmmode {^\circ}\else {$^\circ$}\fi}
\def\porm {\ifmmode {\pm}\else {$\pm$}\fi}
\def\chisqpdf {\ifmmode {\chi^2_{\rm pdf}}\else {$\chi^2_{\rm pdf}$}\fi}
\def\chisq    {\ifmmode {\chi^2}\else {$\chi^2$}\fi}
\def\Msun {M$_\odot$}

\def\etal {et al.~}
\def\eg   {e.g.~}
\def\ie   {i.e.~}

\def\d    {\ifmmode {{\rlap{.}}^\circ}\else {${\rlap{.}}^\circ$}\fi}
\def\s    {\ifmmode {{\rlap{.}}^s}\else {${\rlap{.}}^s$}\fi}
\def\as   {\ifmmode {{\rlap{.}}^{''}}\else {${\rlap{.}}^{''}$}\fi}

\def\pa    {\ifmmode {\psi} \else {$\psi$}\fi}

\def\vlsr  {\ifmmode {v_{\rm LSR}}\else {$v_{\rm LSR}$}\fi}
\def\vlsrr {\ifmmode {v^r_{\rm LSR}}\else {$v^r_{\rm LSR}$}\fi}
\def\vhelio{\ifmmode {v_{Helio}}\else {$v_{Helio}$}\fi}

\def\ura   {\ifmmode {\mu_\alpha}\else {$\mu_\alpha$}\fi}
\def\udec  {\ifmmode {\mu_\delta}\else {$\mu_\delta$}\fi}

\def\ul    {\ifmmode {\mu_l}\else {$\mu_l$}\fi}
\def\ub    {\ifmmode {\mu_b}\else {$\mu_b$}\fi}

\def\uml   {\ifmmode {v_{gr}}\else {$v_{gr}$}\fi}
\def\umb   {\ifmmode {v_b}\else {$v_b$}\fi}
\def\vsrad {\ifmmode {v_{rad}}\else {$v_{rad}$}\fi}

\def\upl   {\ifmmode {v^p_{gr}}\else {$v^p_{gr}$}\fi}
\def\upb   {\ifmmode {v^p_b}\else {$v^p_b$}\fi}
\def\vprad {\ifmmode {v^p_{rad}}\else {$v^p_{rad}$}\fi}

\def\Vo    {\ifmmode {V^{Std}_\odot}\else {$V^{Std}_\odot$}\fi}
\def\Uo    {\ifmmode {U^{Std}_\odot}\else {$U^{Std}_\odot$}\fi}
\def\Wo    {\ifmmode {W^{Std}_\odot}\else {$W^{Std}_\odot$}\fi}
\def\VH    {\ifmmode {V^H_\odot}\else {$V^H_\odot$}\fi}
\def\UH    {\ifmmode {U^H_\odot}\else {$U^H_\odot$}\fi}
\def\WH    {\ifmmode {W^H_\odot}\else {$W^H_\odot$}\fi}
\def\V     {\ifmmode {V_\odot}\else {$V_\odot$}\fi}
\def\U     {\ifmmode {U_\odot}\else {$U_\odot$}\fi}
\def\W     {\ifmmode {W_\odot}\else {$W_\odot$}\fi}

\def\Vs    {\ifmmode {V_s}\else {$V_s$}\fi}
\def\Us    {\ifmmode {U_s}\else {$U_s$}\fi}
\def\Ws    {\ifmmode {W_s}\else {$W_s$}\fi}

\def\Vsbar {\ifmmode {\overline{V_s}}\else {$\overline{V_s}$}\fi}
\def\Usbar {\ifmmode {\overline{U_s}}\else {$\overline{U_s}$}\fi}
\def\Wsbar {\ifmmode {\overline{W_s}}\else {$\overline{W_s}$}\fi}

\def\pars  {\ifmmode{\pi_s}\else{$\pi_s$}\fi}

\def\Ts    {\ifmmode{\Theta_s}\else{$\Theta_s$}\fi}
\def\Tdot  {\ifmmode{d\Theta\over dR}\else{$d\Theta\over dR$}\fi}

\def\Rp    {\ifmmode{R_p}\else{$R_p$}\fi}

\def\To    {\ifmmode{\Theta_0}\else{$\Theta_0$}\fi}
\def\Ro    {\ifmmode{R_0}\else{$R_0$}\fi}

\slugcomment{2009 April 1}
\shorttitle{Galactic Parameters and Structure} 
\shortauthors{Reid \etal}

\begin{document}

\title{Trigonometric Parallaxes of Massive Star Forming Regions: VI.\\ 
   Galactic Structure, Fundamental Parameters and Non-Circular Motions}

\author{M. J. Reid\altaffilmark{1}, K. M. Menten\altaffilmark{2}, 
        X. W. Zheng\altaffilmark{3}, A. Brunthaler\altaffilmark{2},  
        L. Moscadelli\altaffilmark{4}, Y. Xu\altaffilmark{2,5}
        B. Zhang\altaffilmark{3}, M. Sato\altaffilmark{1,6},
        M. Honma\altaffilmark{6}, T. Hirota\altaffilmark{6}, K. Hachisuka\altaffilmark{7},
        Y. K. Choi\altaffilmark{2}, G. A. Moellenbrock\altaffilmark{8}, 
       \& A. Bartkiewicz\altaffilmark{9}
       }

\altaffiltext{1}{Harvard-Smithsonian Center for
   Astrophysics, 60 Garden Street, Cambridge, MA 02138, USA}
\altaffiltext{2}{Max-Planck-Institut f\"ur Radioastronomie, 
   Auf dem H\"ugel 69, 53121 Bonn, Germany}
\altaffiltext{3}{Department of Astronomy, Nanjing University
   Nanjing 210093, China} 
\altaffiltext{4}{Arcetri Observatory, Firenze, Italy}
\altaffiltext{5}{Purple Mountain Observatory, Chinese Academy of
   Sciences, Nanjing 210008, China}
\altaffiltext{6}{VERA Project, National Astronomical Observatory, Tokyo 181 8588, Japan}
\altaffiltext{7}{Shanghai Astronomical Observatory, 80 Nandan Rd., Shanghai, China}
\altaffiltext{8}{National Radio Astronomy Observatory, Socorro, NM, USA}
\altaffiltext{9}{Torun Centre for Astronomy, Nicolaus Copernicus University, 
   Gagarina 11, 87-100 Torun, Poland}

\begin{abstract}
We are using the Very Long Baseline Array and the Japanese VLBI
Exploration of Radio Astronomy project
to measure trigonometric parallaxes and proper motions of masers found in
high-mass star-forming regions across the Milky Way.  Early results from 18 sources 
locate several spiral arms.  The Perseus spiral arm has a pitch angle 
of $16\deg\pm3\deg$, which favors four rather than two spiral arms for the Galaxy.  
Combining positions, distances, proper motions, and radial velocities yields complete 
3-dimensional kinematic information.  We find that star forming regions 
on average are orbiting the Galaxy $\approx15$~\kms\ slower than expected
for circular orbits.  By fitting the measurements to a model of the
Galaxy, we estimate the distance to the Galactic center $\Ro=8.4\pm0.6$~kpc
and a circular rotation speed $\To=254\pm16$~\kms.  The ratio $\To/\Ro$ can be
determined to higher accuracy than either parameter individually, and we find
it to be $30.3\pm0.9$~\kmsperkpc, in good agreement with the angular rotation
rate determined from the proper motion of \SgrA.  The data favor a rotation curve for
the Galaxy that is nearly flat or slightly rising with Galactocentric distance.
Kinematic distances are generally too large, sometimes by factors greater than two;
they can be brought into better agreement with the trigonometric parallaxes 
by increasing $\To/\Ro$ from the IAU recommended value of $25.9$~\kmsperkpc\ to a
value near $30$~\kmsperkpc.  We offer a ``revised'' prescription for calculating
kinematic distances and their uncertainties, as well as a new approach for defining
Galactic coordinates.  Finally, our estimates of \To\ and $\To/\Ro$,
when coupled with direct estimates of \Ro, provide evidence that the
rotation curve of the Milky Way is similar to that of the Andromeda galaxy,
suggesting that the dark matter halos of these two dominant Local Group galaxy
are comparably massive.
\end{abstract}


\section{Introduction}

The Milky Way is known to possess spiral structure.  However, revealing
the nature of this structure has proven to be elusive for decades.
The \citet{Georgelin:76} study of HII regions produced what has been generally
considered the ``standard model'' for the spiral structure of the Galaxy.  
However, after decades of study there is little agreement on this structure.  
Indeed, we do not really know the number of spiral arms
\citep{Simonson:76,Cohen:80,Bash:81,Vallee:95,Drimmel:00,Russeil:03} 
or how tightly wound is their pattern.  
The primary reason for the difficulty is the lack of accurate 
distance measurements throughout the Galaxy.  Photometric distances are
prone to calibration problems, which become especially severe
when looking through copious dust to distant objects in the plane 
of the Galaxy.  Thus, most attempts to map the Galaxy rely on radio frequency 
observations and kinematic distances, which involve matching source Doppler 
shifts with those expected from a model of Galactic rotation.  
However, because of distance ambiguities in the first and fourth quadrants 
(where most of the spiral arms are found) and the existence of sizeable 
non-circular motions, kinematic distances can be highly uncertain 
\citep{Burton:74,Liszt:81,Gomez:06}.

We are measuring trigonometric parallaxes and proper motions of sources of 
maser emission associated with high-mass star forming regions (HMSFRs), 
using the National Radio Astronomy Observatory's
\footnote{The National Radio Astronomy Observatory is a facility of the 
National Science Foundation operated under cooperative agreement by 
Associated Universities, Inc.} Very Long Baseline Array (VLBA)
and the Japanese VLBI Exploration of Radio Astronomy (VERA) project.  
The great advantage of trigonometric parallaxes 
and proper motions is that one determines source distances directly and 
geometrically, with no assumptions about luminosity, extinction,
metallicity, crowding, etc.  Also, from the same measurements,
one determines proper motions, and if the time sampling is optimal
there is little if any correlation between the parallax and 
proper motion estimates.  Thus, the magnitude of the proper motion
does not affect the parallax accuracy.  Combining all of the observational data
yields the full 3-dimensional locations and velocity vectors of the sources.

Results for 12 GHz methanol (\meth) masers toward 10 HMSFRs, carried out with 
the VLBA (program BR100), are reported in the first five papers in this series 
\citep{Reid:09,Moscadelli:09,Xu:09,Zhang:09,Brunthaler:09},
hereafter Papers I through V, respectively.  Eight other sources with \hho\ or
SiO masers have been measured with VERA \citep{Honma:07,Hirota:07,Choi:08,Sato:08} 
and with methanol, \hho\ or continuum emission with the VLBA 
\citep{Hachi:06,Menten:07,Moellenbrock:07,Bartkiewicz:08,Hachi:09}.
In this paper, we collect these parallaxes and proper 
motions in order to study the spiral structure  of the Galaxy. 
Combining positions, distances, Doppler shifts, and proper motions, allows us not 
only to locate the HMSFRs that harbor the target maser sources in 3-dimensions, 
but also to determine their 3-dimensional space motions.  
In \S\ref{sect:spiral_structure} we map the locations of the HMSFRs and measure the 
pitch angles of some spiral arms.  
In \S\ref{sect:gal_dynamics} we use the full 3-dimensional spatial and kinematic 
information to examine the non-circular (peculiar) motions of these star forming regions.
We also fit the data with a model of the Galaxy and estimate the distance from 
the Sun to the Galactic center (\Ro) and the circular orbital speed at the Sun (\To). 
The nature of the rotation curve and its effect on estimates of \Ro\ and \To\
is also discussed.  
In \S\ref{sect:k_dist} we compare kinematic distances with those determined by 
trigonometric parallax and offer a new prescription to improve such distance estimates.
In \S\ref{sect:gal_coordinates} we discuss limitations of the current definition 
of Galactic coordinates and suggest a new system based partly on dynamical
information.  Finally, we discuss the broader implications of our results in 
\S\ref{sect:discussion}.

\section{Galactic Spiral Structure} \label{sect:spiral_structure}

\begin{figure}
\epsscale{1.0} 
\includegraphics{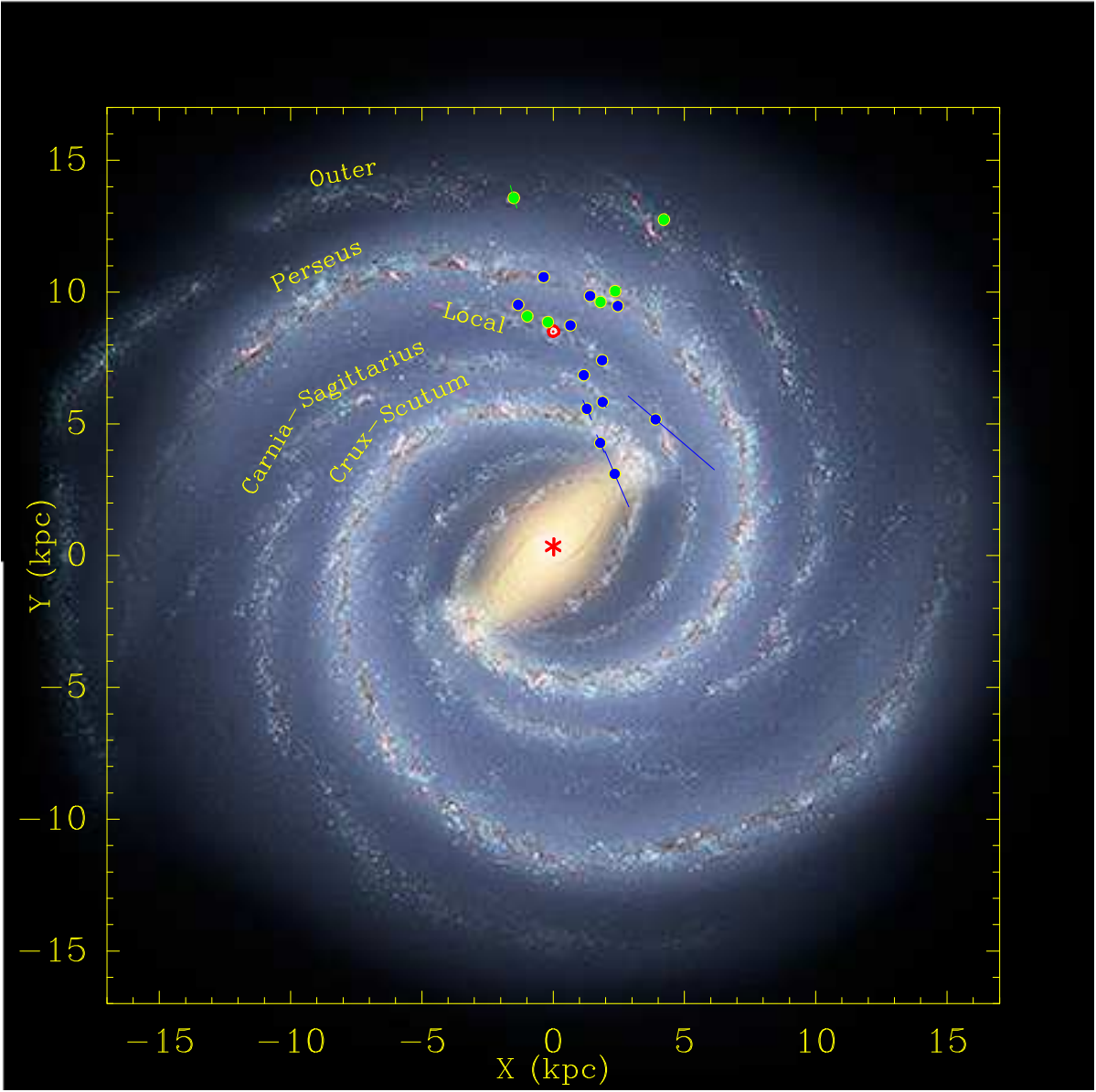}
\caption{\small
Locations of high-mass star forming regions for which 
trigonometric parallaxes have been measured.  Parallaxes from 
12 GHz methanol masers are indicated with {\it dark blue dots} and those
from \hho\ and SiO masers or continuum emission (Orion) are indicated 
with {\it light green dots}.
Distance error bars are indicated, but most are smaller than the dots.
The Galactic center ({\it red asterisk}) is at (0,0) and the Sun 
({\it red Sun symbol}) at (0,8.5).
The background is an artist's conception of Milky Way  
(R. Hurt: NASA/JPL-Caltech/SSC) viewed from the north Galactic pole from which the
Galaxy rotates clockwise.  The artist's image has been scaled to place 
the HMSFRs in the spiral arms, some of which are labeled.}
\label{fig:parallaxes}
\end{figure}

Table~\ref{table:parallaxes} summarizes the parallax and proper motions of 
18 regions of high-mass star formation measured with VLBI techniques.
The locations of these star forming regions in the Galaxy are shown
in Figure~\ref{fig:parallaxes}, superposed on an artist's conception of the Milky Way.   
Distance errors are indicated with error bars ($1\sigma$), but for most sources the 
error bars are smaller than the dots.

\begin{deluxetable}{lrrrrrrl}
\tablecolumns{8} \tablewidth{0pc}
\tablecaption{Parallaxes \& Proper Motions of High-mass Star Forming Regions}
\tablehead {
  \colhead{Source} & \colhead{$\ell$} & \colhead{$b$} &
  \colhead{Parallax} & \colhead{$\mu_x$} & \colhead{$\mu_y$} &
  \colhead{\vlsr}    & \colhead{Ref.}
\\
  \colhead{}      & \colhead{(deg)} & \colhead{(deg)} &
  \colhead{(mas)} & \colhead{(\masy)} & \colhead{(\masy)} &
  \colhead{(\kms)}& \colhead{}         
           }
\startdata
G~23.0$-$0.4...& 23.01 &$-0.41$ &$0.218\pm0.017$ &$-1.72\pm0.04$ &$-4.12\pm0.30$ &$+81\pm3$ & V\\
G~23.4$-$0.2...& 23.44 &$-0.18$ &$0.170\pm0.032$ &$-1.93\pm0.10$ &$-4.11\pm0.07$ &$+97\pm3$ & V\\
G~23.6$-$0.1...& 23.66 &$-0.13$ &$0.313\pm0.039$ &$-1.32\pm0.02$ &$-2.96\pm0.03$ &$+83\pm3$ & 1 \\
G~35.2$-$0.7...& 35.20 &$-0.74$ &$0.456\pm0.045$ &$-0.18\pm0.06$ &$-3.63\pm0.11$ &$+28\pm3$ & IV\\
G~35.2$-$1.7...& 35.20 &$-1.74$ &$0.306\pm0.045$ &$-0.71\pm0.05$ &$-3.61\pm0.17$ &$+42\pm3$ & IV\\
W~51~IRS~2..   & 49.49 &$-0.37$ &$0.195\pm0.071$ &$-2.49\pm0.08$ &$-5.51\pm0.11$ &$+56\pm3$ & III\\
G~59.7$+$0.1...& 59.78 &$+0.06$ &$0.463\pm0.020$ &$-1.65\pm0.03$ &$-5.12\pm0.08$ &$+27\pm3$ & III\\
Cep~A...........&109.87&$+2.11$ &$1.430\pm0.080$ &$+0.50\pm1.10$ &$-3.70\pm0.20$ &$-10\pm5$ & II\\
NGC~7538....   &111.54 &$+0.78$ &$0.378\pm0.017$ &$-2.45\pm0.03$ &$-2.44\pm0.06$ &$-57\pm3$ & II\\
IRAS~00420..   &122.02 &$-7.07$ &$0.470\pm0.020$ &$-1.99\pm0.07$ &$-1.62\pm0.05$ &$-44\pm5$ & 2 \\
NGC~281......  &123.07 &$-6.31$ &$0.355\pm0.030$ &$-2.63\pm0.05$ &$-1.86\pm0.08$ &$-31\pm5$ & 3\\
W3(OH).......  &133.95 &$+1.06$ &$0.512\pm0.010$ &$-1.20\pm0.20$ &$-0.15\pm0.20$ &$-45\pm3$ & 4\\
WB~89-437...   &135.28 &$+2.80$ &$0.167\pm0.006$ &$-1.27\pm0.50$ &$+0.82\pm0.05$ &$-72\pm3$ & 5\\
S~252............&188.95&$+0.89$&$0.476\pm0.006$ &$+0.02\pm0.01$ &$-2.02\pm0.04$ &$+11\pm3$ & I\\
S~269............&196.45&$-1.68$&$0.189\pm0.016$ &$-0.42\pm0.02$ &$-0.12\pm0.08$ &$+20\pm3$ & 6\\
Orion............&209.01&$-19.38$&$2.425\pm0.035$&$+3.30\pm1.00$ &$+0.10\pm1.00$ &$+10\pm5$ & 7\\
G~232.6$+$1.0..&232.62 &$+1.00$ &$0.596\pm0.035$ &$-2.17\pm0.06$ &$+2.09\pm0.46$ &$+23\pm3$ & I\\
VY~CMa.......  &239.35 &$-5.06$ &$0.876\pm0.076$ &$-3.24\pm0.16$ &$+2.06\pm0.60$ &$+18\pm3$ & 8\\
\enddata
\tablecomments {
 {\footnotesize
   Columns 2 and 3 give Galactic longitude and latitude, respectively.  
   Columns 5 and 6 are proper motions in the eastward ($\mu_x=\ura \cos{\delta}$) 
   and northward directions ($\mu_y=\udec$), respectively.  
   Column 7 lists  Local Standard of Rest velocity components; these can be converted to a 
   heliocentric frame as described in the Appendix.
   References are I: \citet{Reid:09}; II: \citet{Moscadelli:09}; 
   III: \citet{Xu:09}; IV: \citet{Zhang:09}; V: \citet{Brunthaler:09};
   1: \citet{Bartkiewicz:08}; 2: \citet{Moellenbrock:07}; 3: \citet{Sato:08}; 
   4: \citet{Xu:06,Hachi:06}; 5: \citet{Hachi:09}; 6: \citet{Honma:07};
   7: \citet{Hirota:07,Menten:07}; 8: \citet{Choi:08}.     
   The calculations in this paper use an early parallax and proper motion estimate for 
   IRAS 00420+5530 cited above; the values reported more recently by
   \citet{Moellenbrock:09} are slightly different but would not substantively change the 
   results presented here.
  }
}
\label{table:parallaxes}
\end{deluxetable}

\subsection{Spiral Arms} \label{sect:spiral_arms}

The HMSFRs with parallaxes locate several spiral arms.
The three sources closest to the Galactic center (G~23.0$-$0.4, G~23.4$-$0.2,
and G~23.6$-$0.1) appear to be members of the Crux-Scutum or possibly
the Norma or the 3-kpc arm.   However, the parallax uncertainties for these
distant, low-declination sources are currently not adequate to clearly 
distinguish among these arms, especially in the crowded region
where the Galactic bar (see \citet{Blitz:91b} and references therein)
ends and the arms begin \citep{Benjamin:05,Dame:08}.

Three sources (G~35.2$-$0.7, G~35.2$-$1.7 and W~51~IRS~2) are probably in the 
Carina-Sagittarius arm, whose distance from the Sun is 2.5~kpc at Galactic
longitude, $\ell$, of $\approx 35\deg$.  

Five sources (S~252, W3(OH), IRAS~00420+5530, NGC~281, and NGC~7538)
clearly trace a portion of the Perseus arm, which is located between
distances of 2.10~kpc at $\ell=189\deg$ (S~252) and 2.64~kpc at $\ell=112\deg$ 
(NGC~7538).  NGC~281 is slightly offset from the other sources in the Perseus 
arm and is believed to be associated with an expanding super-bubble \citep{Sato:08}.
As such, it may not accurately trace spiral structure.

Two sources, (S~269 and WB~89-437), measured by \citet{Honma:07} with the VERA 
array and \citet{Hachi:09} with the VLBA, lie beyond the Perseus arm and begin 
to trace an Outer (Cygnus) arm at a distance from the Sun of 5.3 kpc at 
$\ell=196\deg$ for S~269 to 5.9~kpc at $\ell=135\deg$ for WB~89-437.

The remaining five sources (G~232.6$+$1.0, VY~CMa, the Orion Nebula, Cep~A 
and G59.7$+$0.1) trace the Local (Orion) ``arm,'' which appears to be a spur 
between the Carina--Sagittarius and Perseus arms.  
The Sun is in or near this spur, and we can trace it between
G~59.7$+$0.1 near the Carina--Sagittarius arm at $\ell=60\deg$ and
G~232.6$+$1.0 near the Perseus arm at $\ell=233\deg$.

\subsection{Pitch Angles} \label{sect:pitch_angles}

\begin{figure}[htp]
\epsscale{0.8} 
\plotone{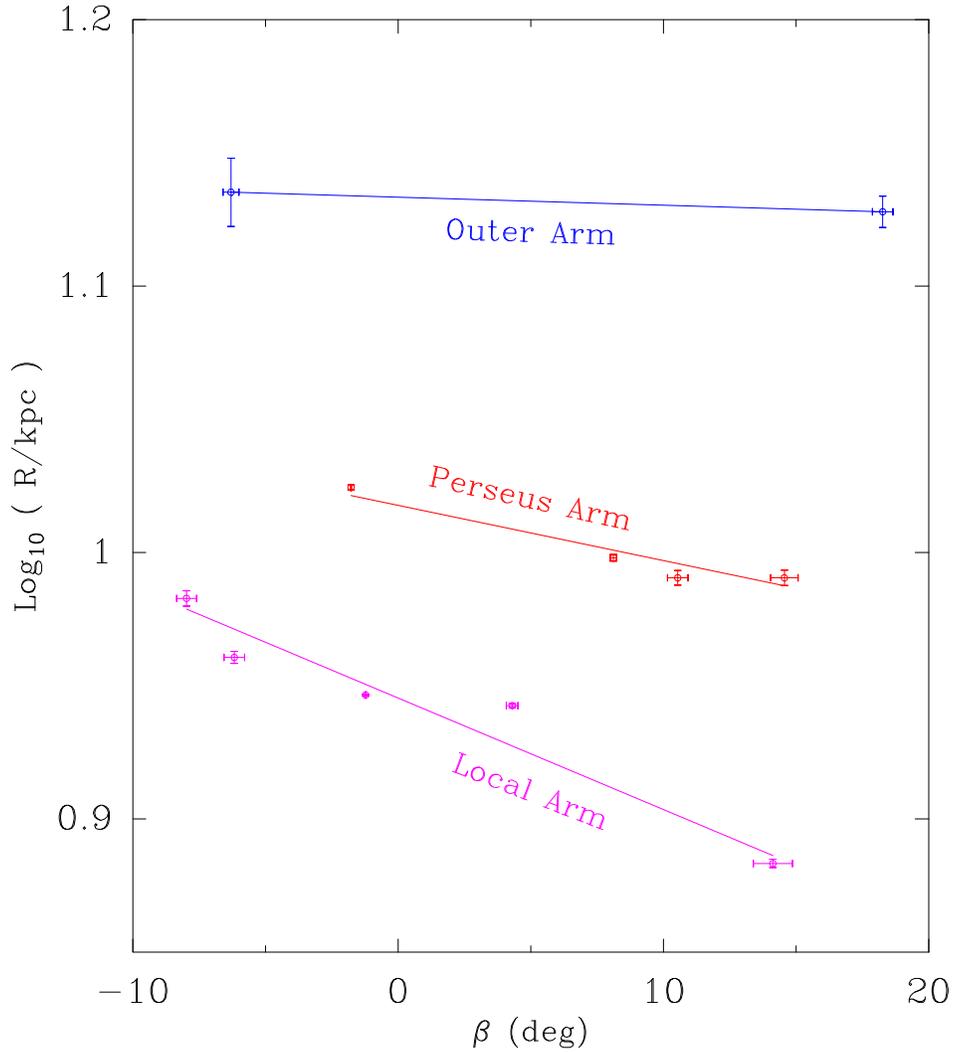}  
\caption{\small
Spiral arm pitch angles. The logarithm of Galactocentric radius,
$R$, (in kpc units) is plotted against Galactocentric longitude ($\beta$).
Data based on trigonometric parallaxes for sources that can be confidently
assigned to a spiral arm are shown along with $1\sigma$ uncertainties.
Positional variations of star forming regions within an arm are clearly 
greater than the parallax uncertainties.   Unweighted fits of straight
lines to the data are shown with solid lines.  Pitch angles are proportional 
to the negative of the arctangent of the line slopes.
        }  
\label{fig:pitch_angles}
\end{figure}

Spiral arm pitch angles can be estimated when two or more
sources can be confidently identified as members of a single arm.  
The pitch angle, \pa, of an arm segment can be defined by constructing a
line segment between sources in the same section of the arm.
Next, construct a line tangential to a Galactocentric circle that
passes through the midpoint of this segment and determine the angle 
between these two lines.  An ideal log-periodic spiral arm can be defined by 
the equation $$\ln{(R/R_{ref})} = -(\beta - \beta_{ref}) \tan{\pa}~~,$$
where $R$ is the Galactocentric radius at a Galactocentric
longitude $\beta$ (defined as 0 toward the Sun and increasing 
with Galactic longitude) for an arm with a reference radius $R_{ref}$ at
$\beta_{ref}$.  In Figure~\ref{fig:pitch_angles}, we plot
$\log{(R/1~{\rm kpc})}$ versus $\beta$ for the three arms where we can clearly
identify two or more HMSFRs.  In such a plot, spiral arm sections appear
as straight lines.  Some of the data deviate from fitted lines by considerably
more than the parallax errors, as expected for variations of
the locations of star forming regions within an arm whose width is $\sim100$~pc.
Thus we used unweighted straight line fits to estimate spiral arm pitch angles.

The Perseus arm sources (excluding NGC~281) indicate a pitch angle of 
$16\d5\porm3\d1$ between Galactic longitude 112\deg\ and 189\deg.   
This is in the upper half of pitch angle estimates of 5\deg\ to 21\deg\ 
for spiral arms in the Galaxy collected by \citet{Vallee:95}.  
The five HMSFRs with parallaxes that trace the Local ``arm'' indicate
a mean pitch angle of $27\d8\porm4\d7$.  However, the Local arm is probably 
not a global spiral arm; instead it appears to be a short segment or
spur between the Carina--Sagittarius and Perseus arms.  

Two sources (S~269 \& WB89--437) appear to be part of the Outer arm and formally
yield a pitch angle of $2\d3$.  This suggests that the
Outer arm might have a smaller pitch angle than the Perseus arm.
This may be of significance, but with only two sources, 
more parallaxes are needed before reaching any firm conclusions.

For other spiral arms, we have too few parallaxes to reliably 
determine pitch angles.  The sources that are possible members
of the Carina--Sagittarius arm (G~34.2$-$0.7, G~34.2$-$1.7 \& W~51~IRS~2)
would formally give a wide range of pitch angles.  However, because 
one or more sources might be associated with the Local arm, we cannot
reliably estimate the pitch angle of the Carina--Sagittarius arm at this time.

\section{Galactic Dynamics} \label{sect:gal_dynamics}

Given measurements of position, parallax, proper motion and Doppler shift,
one has complete three-dimensional location and velocity vectors relative 
to the Sun.  One can then construct a model of the Milky Way and adjust the 
model parameters to best match the data.  We model the Milky Way as a 
disk rotating with speed $\Theta(R)=\To+\Tdot~(R-\Ro)$, 
where \Ro\ is the distance from the Sun to the Galactic center. 
We started the fitting process by assuming a flat rotation curve (\ie $\Tdot=0$).
Later, we relaxed this assumption and solved for $\Tdot$, followed by an investigation of
other forms of the Galactic rotation curve.  
Since all measured motions are relative to the Sun, we need to remove 
the peculiar (non-circular) motion of the Sun, which is parameterized by 
\U\ toward the Galactic center, \V\ in the direction of Galactic rotation, and 
\W\ towards the north Galactic pole (NGP).  
Table~\ref{table:model} summarizes these and other parameters.

\begin{deluxetable}{ll}
\tablecolumns{2} \tablewidth{0pc} 
\tablecaption{Galaxy Model Parameter Definitions}
\tablehead {
  \colhead{Parameter} & \colhead{Definition} 
            }
\startdata
\Ro ......  & Distance of Sun from GC \\
\To ......  & Rotation Speed of Galaxy at \Ro\\
\Tdot ..... & Derivative of $\Theta$ with $R$: $\Theta(R)=\To+\Tdot~(R-\Ro)$\\
\U .......  & Solar motion toward GC  \\
\V .......  & Solar motion in direction of Galactic rotation\\
\W .......  & Solar motion toward NGP \\
\Usbar ......  & Average source peculiar motion toward GC  \\
\Vsbar ......  & Average source peculiar motion in direction of Galactic rotation\\
\Wsbar ......  & Average source peculiar motion toward NGP \\
\enddata
\tablecomments {\footnotesize
GC is the Galactic Center and NGP is the North Galactic Pole.  
The average source peculiar motions (\Usbar,\Vsbar,\Wsbar) are defined at the location of the 
source and are rotated with respect to the solar motion (\U,\V,\W) by the 
Galactocentric longitude, $\beta$, of the source (see Appendix Figure~\ref{fig:definitions}).
We solve for the magnitude of each component, but the orientation of the vector
for each source depends on location in the Galaxy.
               } 
\label{table:model}
\end{deluxetable}

We adjusted the Galactic parameters so as to best match the data to the 
spatial-kinematic model in a least-squares sense.  For each source, we treated 
the measured parallax (\pars), two components of proper motion ($\mu_x,\mu_y$), 
and the heliocentric velocity (\vhelio) as data.
The observed source coordinates are known to extremely high
accuracy and were treated as independent variables.
The model is a smoothly rotating galaxy given by the parameters listed in
Table~\ref{table:model}.  Specifically, the model parallax is calculated from
a kinematic distance, based on the observed Doppler shift.  The three-dimensions
of motion relative to the Sun (proper motion and heliocentric Doppler shift)
are calculated from the source location, taking into account the size and rotation
curve of the galaxy model, and the solar and source peculiar motions.  
We adopt the Hipparcos determination of solar motion \citep{Dehnen:98}
as definitive and generally did not vary these parameters.  However, in one least-squares
fit, we solved for these parameters for illustrative purposes in order to compare 
Solar Motion results from Hipparcos stars and our HMSFRs. 

Our choice of weights for the data in the least-squares fitting process requires some
comment.  While the heliocentric velocity of any maser spot can be measured
with very high accuracy, it may not exactly reflect the motion of
the HMSFR.  The internal motions of methanol masers are generally small and
cause uncertainty of $\approx3$~\kms\ \citep{Moscadelli:02}, whereas \hho\ masers
can be associated with fast outflow and, if not accurately modeled, can lead to
larger uncertainty in the motion of the exciting star.  In addition, the
Virial motion of an individual massive star (associated with the masers) 
with respect to the entire HMSFR is likely to be $\approx7$~\kms\ per coordinate 
(\eg  for a region of mass of $\sim3\times10^4$~\Msun\ and radius of $\sim1$~pc).
Therefore, we allow for a deviation of the measured motion from the center of mass 
of its associated HMSFR by adding an uncertainty of $\sigma_{Vir}=7$~\kms\ 
in quadrature with the internal motion estimates (between 3 and 5~\kms).
Specifically, the variance weights for the \vlsr\ data, $w(\vlsr)$, are calculated
from $w(\vlsr) = 1/(\sigma_\vlsr^2 + \sigma_{Vir}^2)$.

Since the parallax data is compared to a kinematic model, we considered both the 
parallax measurement uncertainty and a modeling uncertainty for the kinematic 
distance, $\sigma_{kd}$, owing to the total uncertainty in the heliocentric 
velocity of the associated HMSFR.  These two components were added in quadrature 
when calculating the weights: 
$w(\pars) = 1/(\sigma_\pi^2 + \sigma_{kd}^2/d_s^4)$, where $d_s=1/\pars$.
Similarly, the proper motion weights allowed for both 
measurement uncertainties and the possible deviation of the measured maser 
motions from the center of mass of the HMSFR.  The latter term was set by the
uncertainty in the heliocentric velocity divided by the distance.  
Thus, for either proper motion component
$w(\mu) = 1/(\sigma_\mu^2 + \sigma_{Vir}^2/d_s^2)$.

\subsection{Galactic 3-D Motions} \label{sect:3-D}

We first used all 18 sources listed in Table~\ref{table:parallaxes} 
and solved only for the fundamental Galactic parameters, yielding
$\Ro=8.2$~kpc and $\To=265$~\kms\ for a flat rotation curve ($\Tdot=0$)
(see Fit~1 in Table~\ref{table:fits}).  
The \chisq\ value of 263 for 70 degrees of freedom was quite large, and   
the post-fit residuals showed clear systematic deviations, indicating
a deficiency in this 2-parameter model for Galactic dynamics.

\begin{figure}[htp]
\epsscale{1.0} 
\plotone{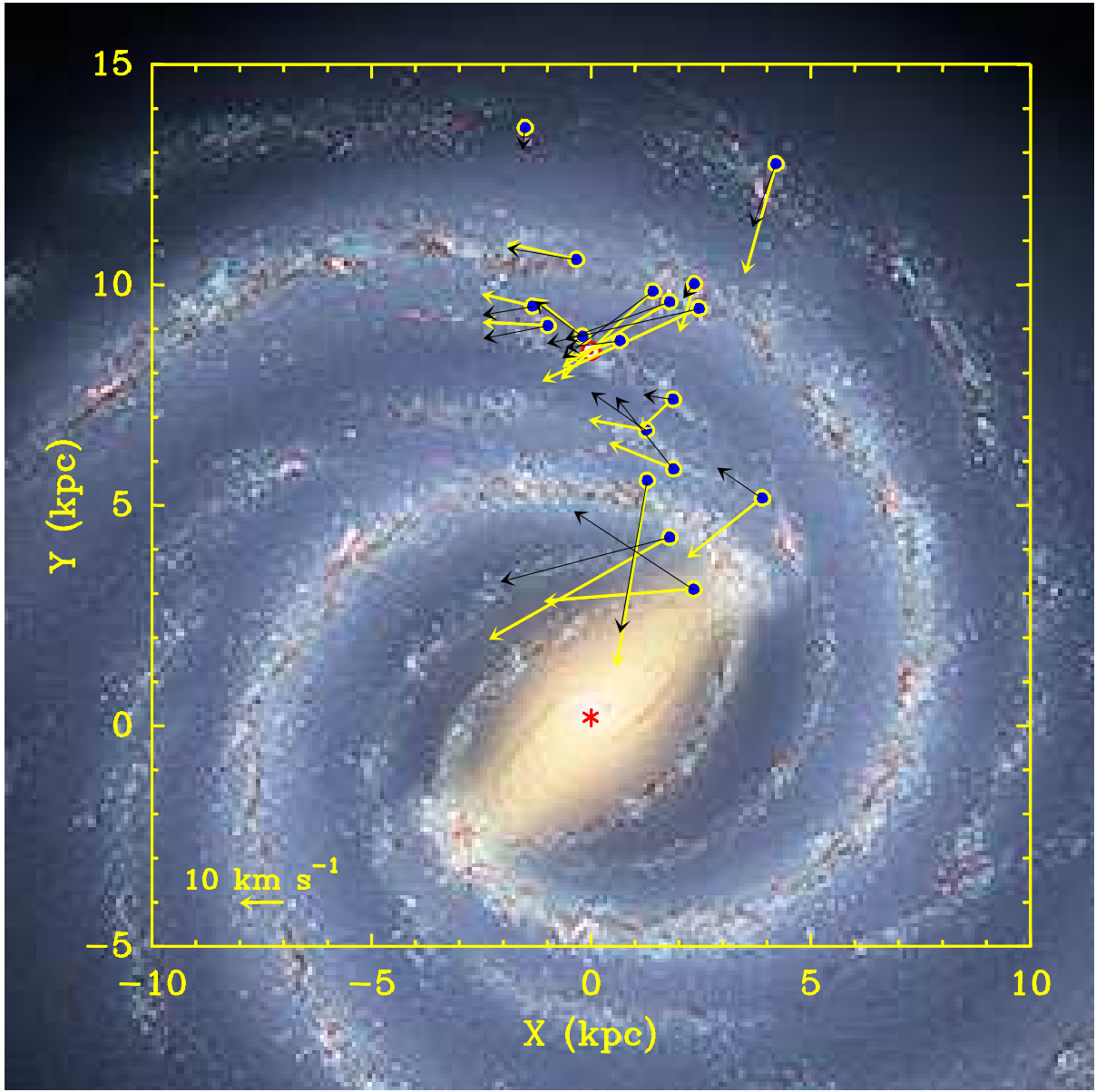}  
\caption{\small
Peculiar motion vectors of high mass star forming regions 
(superposed on an artist conception) projected on the Galactic plane after transforming 
to a reference frame rotating with the Galaxy.  A 10~\kms\ motion scale is in the 
lower left.  The Galaxy is viewed from the north Galactic pole and rotates clockwise.  
The light (yellow) arrows are for IAU standard values of $\Ro=8.5$~kpc and $\To=220$~\kms\ 
and a flat rotation curve, whereas black arrows are for $\To=254$~\kms.  This demonstrates
that the qualitative result that high mass star forming regions orbit the Galaxy 
slower than the Galaxy rotates is not sensitive to the value of \To.
        }  
\label{fig:peculiar_motions}
\end{figure}

Figure~\ref{fig:peculiar_motions} shows the peculiar motions of the HMSFRs
in the Galactic plane by transforming to a reference frame that rotates with the
Galaxy.  Peculiar motions relative to two Galactic rotation models are shown, 
one for $\To=220$~\kms\ (the IAU recommended value) and the other for 
$\To=254$~\kms\ (our best fit value from Section \ref{sect:gal_model} below). 
Both transformations assume $\Ro=8.5$~kpc, a flat rotation curve, and the Hipparcos 
solar motion of \citet{Dehnen:98}.  (The equations used for the transformation are 
documented in the Appendix.)   Sizeable systematic motions are clearly
evident --- almost all sources have a significant component of peculiar
motion {\it counter} to Galactic rotation.
On average these star forming regions orbit the Galaxy $\approx15$~\kms\ slower than
the Galaxy spins.  As is evident from the two sets of peculiar motions, 
this conclusion is insensitive to the values adopted for \To.
Similarly, adopting a more complex rotation curve, \eg the \citet{Clemens:85} curve, 
would not change the qualitative nature of the residuals. 
{\it HMSFRs appear to orbit the Galaxy slower than for circular orbits.} 
This might be explained by star formation triggered by the encounter of 
molecular gas with a shock front associated with a trailing spiral arm
and may help explain the 17~\kms\ dispersion seen in HI data by \cite{Brand:93}.

For the distribution of sources in our sample, the solar motion 
parameters \U\ and \V\ can partially mimic the average source peculiar motions. 
We believe the solar motion parameters determined from
Hipparcos data by \citet{Dehnen:98} are well determined, and they have
been independently confirmed by \citet{Mendez:99}, based on the Southern Proper-Motion 
program data.  However, it is instructive to solve for the solar motion parameters 
with the parallax and proper motion data.  Doing so we find an acceptable fit with 
$\Ro=8.4$~kpc, $\To=242$~\kms, $\U=9$~\kms, $\V=20$~\kms\ and $\W=10$~\kms.
(The~\chisq\ value for this fit was 67.2 for 59 degrees of freedom, which
is somewhat worse than the value of 65.7 found in \S\ref{sect:gal_model}, 
where we adopt the Hipparcos solar motion parameters and solved instead 
for average source peculiar motion components.)

\begin{figure}[htp]
\epsscale{0.7} 
\plotone{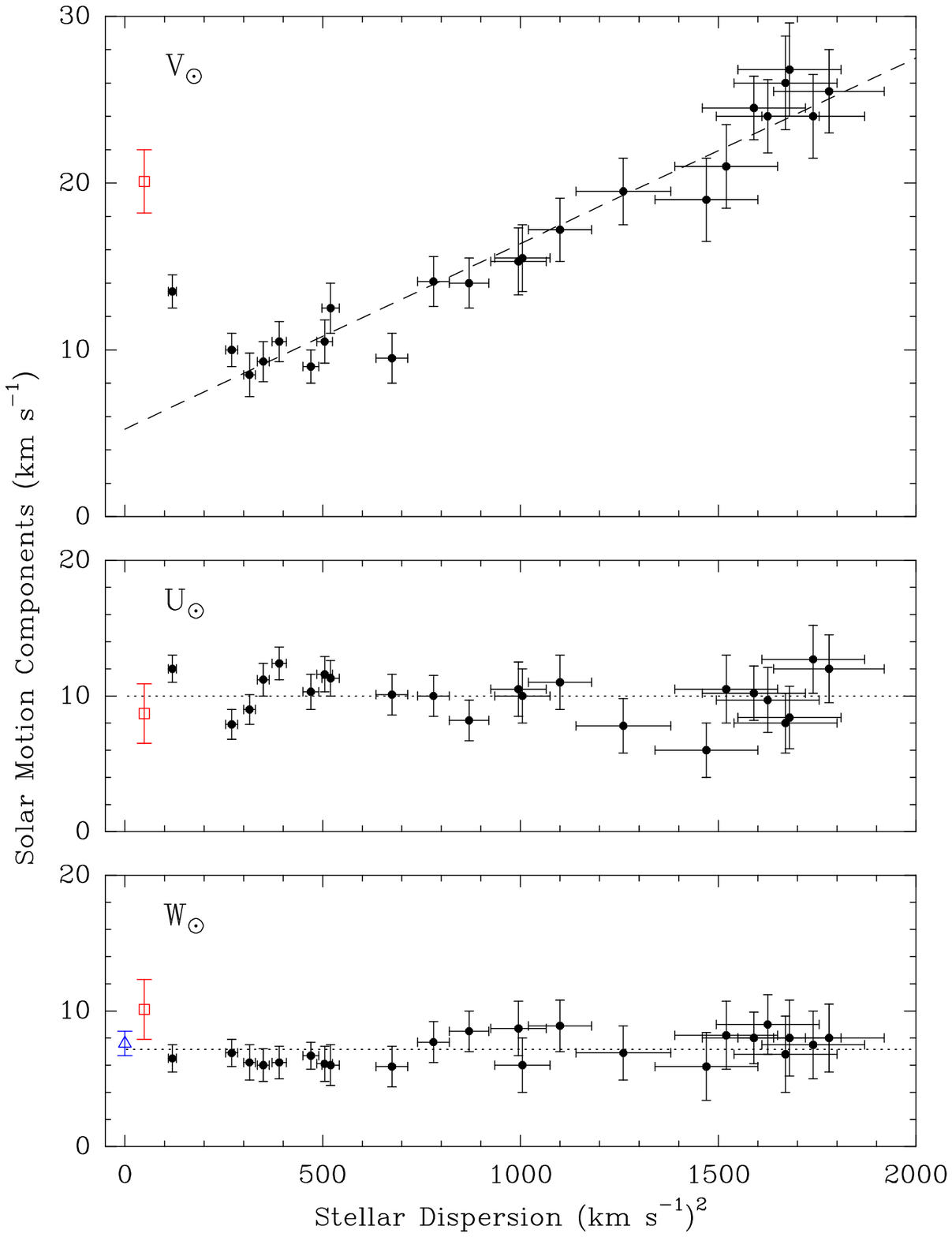} 
\caption{\small
Solar motion components determined from Hipparcos stars 
(\ie the reflex of the average motion of stars) versus stellar
velocity dispersion after \citet{Dehnen:98}.  
{\it Top Panel: } \V\ is the Solar Motion in the direction of Galactic rotation 
(\ie toward $\ell=90\deg$). The ``asymmetric drift'' is shown with the {\it dashed line}.
{\it Middle Panel: } \U\ is the Solar Motion toward the Galactic center. 
{\it Bottom Panel: } \W\ is toward the north Galactic pole.  
Also plotted at $50$ (\kms)$^2$ dispersion with {\it open red squares} are solar motion 
parameters obtained from the parallax and proper motions of star forming regions, and
at zero dispersion with an {\it open triangle} is the \W\ component inferred 
from the proper motion of \SgrA\ by \citet{Reid:04}.
Note the good agreement of the \U\ and \W\ components between Hipparcos and this study.
The large deviation of the \V\ component from the asymmetric drift from this study
is not indicative of large \V\ value, but points to a significant deviation from circular
orbits for very young stars.  
        }
\label{fig:hipparcos}
\end{figure}

In Figure~\ref{fig:hipparcos} we reproduce the Hipparcos solar motion data from 
Figure~4 of \citet{Dehnen:98}.  Their data were binned by stellar colors, plotted
against stellar dispersion and the solar motion components estimated as 
minus the average velocity of all stars in each bin.  
We have also plotted our estimates of the solar motion,
plotted at near-zero ``stellar dispersion'' appropriate for newly formed stars.
Also included in the bottom panel is the value of \W\ determined from the
proper motion of \SgrA\ by \citet{Reid:04}, assuming that the supermassive
black hole is stationary at the Galactic center.
These values for \U\ and \W\ are in good agreement with the Hipparcos results.

The Hipparcos data used to determine \V\ (the solar motion component in the direction 
of Galactic rotation) clearly show the well known ``asymmetric drift,'' which
when extrapolated to zero dispersion should define the Local Standard of Rest (LSR).  
Our value of $\V=20$~\kms, based on HMSFR parallaxes and proper motions, is far above 
the asymmetric drift line, indicating that the HMSFRs as a group are orbiting the 
Galaxy slower than for circular orbits.  Note that the youngest stars in the 
Hipparcos data, plotted at a dispersion of $\approx120$~(\kms)$^2$, show a similar, 
but not as great a departure from the asymmetric drift line.  Evidence that young 
stars lag the LSR orbit has also been found by \citet{Zab:02}. 

Finally, we note that we find no evidence for a global motion of the LSR 
(\ie disagreement with the Hipparcos solar motion) in the direction of the
Galactic center or out of the plane of the Galaxy larger than $6$~\kms\ ($2\sigma$).  
This is contrary to the conclusions of \citet{Kerr:62} and \citet{Blitz:91a}, based 
on an analysis of HI data, that the LSR is moving away from the Galactic center 
at a speed of $>10$~\kms.  

\subsection{Fundamental Galactic Parameters} \label{sect:gal_model}

Since, as shown in \S\ref{sect:3-D}, HMSFRs are orbiting the Galaxy slower than for 
circular orbits, we must allow for such effects when modeling the
Galaxy.   In order to determine the fundamental parameters \Ro\ and \To, we solved for 
three additional parameters, allowing for an average peculiar motion for
all sources with components \Usbar\ toward the Galactic center (as seen by the source),
\Vsbar\ in the local direction of Galactic rotation and \Wsbar\ toward the north 
Galactic pole.  This solution, listed as Fit~2 in Table~\ref{table:fits}, yields 
$\Ro=8.5$~kpc and $\To=264$~\kms\ and peculiar motion components 
of $\Usbar=4$~\kms, $\Vsbar=-16$~\kms\ and $\Wsbar=3$~\kms.   
The residuals show greatly reduced systematic deviations, and the 
\chisq\ value improved significantly to 112 for 67 degrees of freedom, compared to
the solution without the average peculiar motions (Fit~1 in Table~\ref{table:fits}).  

\begin{figure}[htp]
\epsscale{1.0} 
\plotone{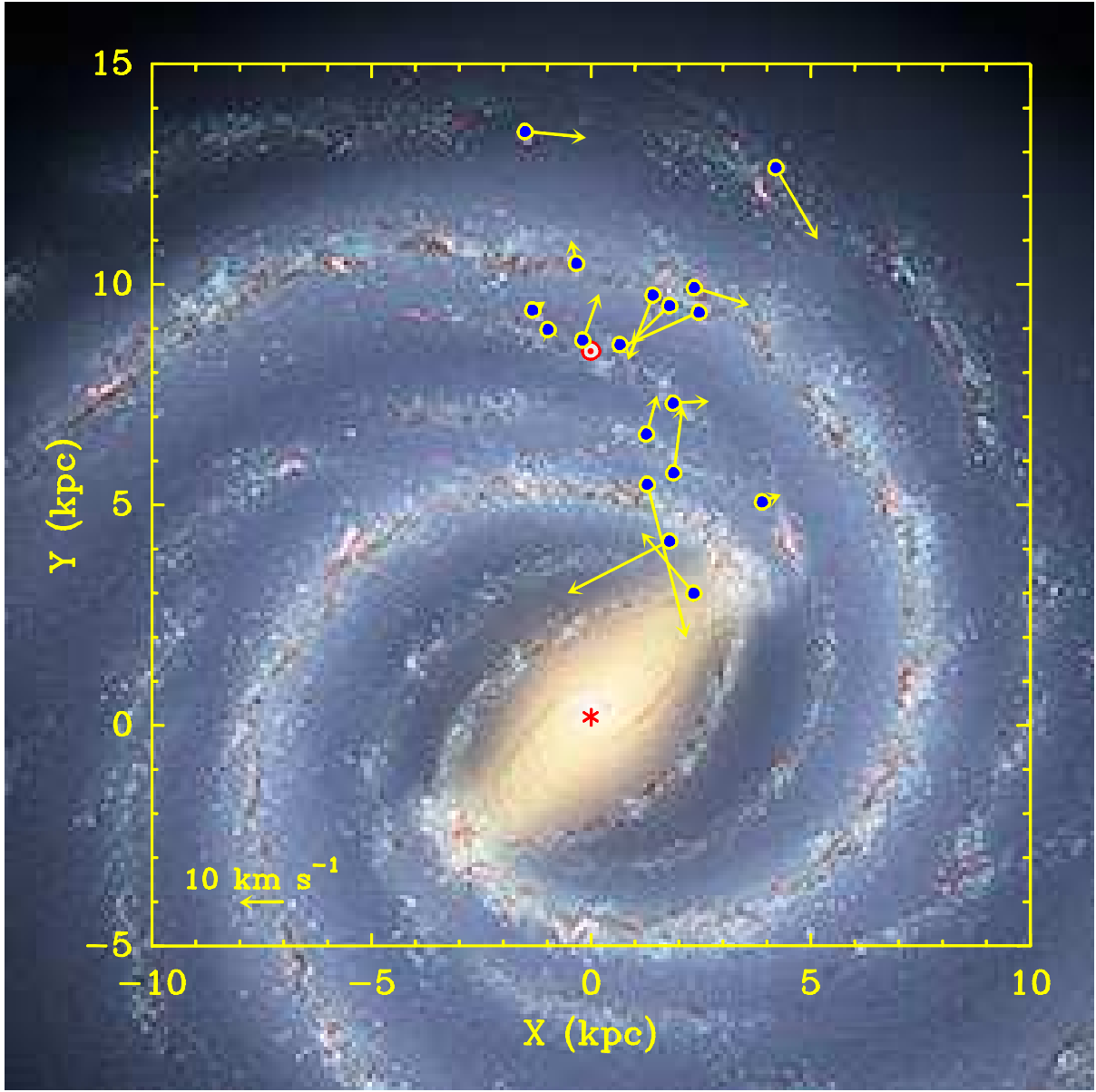}  
\caption{\small
Peculiar motion vectors of high mass star forming regions 
(superposed on an artist conception) after transforming 
to a reference frame rotating with the Galaxy, using best-fit values of 
$\Ro=8.4$~kpc and $\To=254$~\kms\ and removing an average motion of 15~\kms\ counter to 
Galactic rotation and 2~\kms\ toward the Galactic center.  A 10~\kms\ motion scale is 
in the lower left.  The Galaxy is viewed from the north Galactic pole and rotates 
clockwise.  
        }  
\label{fig:best_motions}
\end{figure}

Two sources from the sample, NGC~7538 and G~23.6$-$0.1, displayed post-fit 
residuals significantly greater ($>3\sigma$) than the others.
Removing these sources, we arrive at our ``basic sample'' of 16 HMSFRs.
We repeated the fitting and found $\Ro=8.40\pm0.36$~kpc, $\To=254\pm16$~\kms, 
$\Usbar=2.3\pm2.1$~\kms, $\Vsbar=-14.7\pm1.8$~\kms\ and $\Wsbar=3.0\pm2.1$~\kms\ 
(see Fit~3 in Table~\ref{table:fits}).  The \chisq\ value for this sample was 
considerably improved: 65.7 for 59 degrees of freedom.  The near-zero average 
motion out of the plane of the Galaxy (\Wsbar) is as expected for massive
star forming regions.  The residual motions in the plane of the Galaxy are shown in 
Figure~\ref{fig:best_motions}.  Most of the star 
forming regions have residual velocities consistent with measurement error combined with
expected Virial motions within HMSFRs of $\sim7$~\kms\ per coordinate.  
The most distant sources at low Declination (and low Galactic longitude) have larger 
residual velocities owing to greater parallax and proper motion measurement uncertainty 
and the scaling of proper motions to linear speeds by multiplying by distance.

We feel that this solution provides the best estimates of the parameters for
the current data set, under the assumption of a flat rotation curve.  
In \S\ref{sect:rotation_curves} we show that the estimate of \Ro\ is somewhat 
sensitive to the nature of the rotation curve of the Galaxy, leading to a 
systematic source of uncertainty for \Ro\ of approximately $\pm0.5$~kpc.
Combining the statistical and systematic uncertainties in quadrature, we find
$\Ro=8.4\pm0.6$~kpc.

The correlation coefficient between \Ro\ and \To\ was 0.87, while all others 
were small.  This is expected, since kinematic model distances increase
with \Ro\ and inversely with \To.  Thus, the ratio \To/\Ro, which is the angular rotation
rate of the LSR, is determined to much better accuracy than either parameter separately.
Holding $\Ro=8.50$~kpc\ (the IAU recommended value), we find $\To=257.9\pm7.7$~\kms\ or 
$\To/\Ro=30.3\pm0.9$~\kmsperkpc. There is only a slight dependence of $\To/\Ro$ on the 
value adopted for \Ro.  For example, setting $\Ro=8.00$~kpc, we obtain 
$\To/\Ro=30.0\pm0.9$~\kmsperkpc.  See \S\ref{sect:discussion} for a discussion
of the significance of this result.

As shown in \S\ref{sect:rotation_curves}, while estimates of \To\ change by  
$\pm20$~\kms\ among the fits using different rotation curves, this variation can be 
accounted for mostly through the correlation with \Ro, and, therefore, the least-squares 
fitting process incorporates this correlation in the formal uncertainty estimate.
Thus, we conclude that the formal uncertainty of $\pm16$~\kms\ for \To\ is reasonable
(provided that \Ro\ is within 0.5~kpc of 8.4~kpc).   When \Ro\ is ultimately 
measured with much higher accuracy, \To\ would be even better determined from the well 
determined ratio of $\To/\Ro$.

We also considered the possibility that a large positive value for \Usbar\ 
(toward the Galactic center), as could be expected from spiral density wave theory, 
might inflate the estimate of \To.  Holding $\Usbar=17$~\kms (15~\kms greater than 
our best fit) did not significantly reduce the estimate of \To, but did dramatically 
increase the \chisq\ to 200.1.  Thus, we exclude a large $\Usbar$ value and that 
it could contribute to significant uncertainty in \To.

\begin{deluxetable}{ccccccccccc}
\tablecolumns{11} \tablewidth{0pc} \rotate
\tablecaption{Least-squares Fitting Results}
\tablehead {
  \colhead{Fit} & \colhead{\Ro} & \colhead{\To} &  \colhead{\Tdot}  & 
  \colhead{\Usbar}  & \colhead{\Vsbar} & \colhead{\Wsbar} &
  \colhead{\chisq} & \colhead {DF} &
  \colhead{\To/\Ro} 
\\
  \colhead{}      & \colhead{(kpc)} & \colhead{(\kms)} & \colhead{(\kmsperkpc)} &
  \colhead{(\kms)} & \colhead{(\kms)} & \colhead{(\kms)} & 
  \colhead{}       & \colhead {}     &
  \colhead{(\kmsperkpc)} 
           }
\startdata
 1 & 8.24\porm0.55 & 265\porm26 & 0.0\err   &   0.0\err     & \p0.0\err      & 0.0\err 
   & 263.3  & 70 & 32.4\porm1.3  \\
 2 & 8.50\porm0.44 & 264\porm19 & 0.0\err   &   3.9\porm2.5 &$-$15.9\porm2.1 & 3.1\porm2.5  
   & 111.5  & 67 & 31.1\porm1.1  \\
 \\
 3 & 8.40\porm0.36 & 254\porm16 & 0.0\err   &   2.3\porm2.1 &$-$14.7\porm1.8 & 3.0\porm2.2  
   & \p66.7 & 59 & 30.3\porm0.9  \\
\\
 4 & 9.04\porm0.44 & 287\porm19 &2.3\porm0.9&   1.9\porm2.0 &$-$15.5\porm1.7 & 3.0\porm2.1  
   & \p59.0 & 58 & 31.1\porm0.9  \\
 5 & 8.73\porm0.37 & 272\porm15 &Clemens-10 &   1.7\porm1.9 &$-$12.2\porm1.7 & 3.1\porm1.9  
   & \p52.9 & 59 & 31.0\porm0.8  \\
 6 & 7.88\porm0.30 & 230\porm12 &Clemens-8.5&   2.7\porm2.2 &$-$12.4\porm1.9 & 3.1\porm2.3  
   & \p71.2 & 59 & 29.6\porm1.0  \\
 7 & 8.79\porm0.33 & 275\porm13 &Brand-Blitz&   1.9\porm2.0 &$-$18.9\porm1.8 & 3.0\porm2.1  
   & \p59.0 & 59 & 31.0\porm0.9  \\
\enddata
\tablecomments{\footnotesize
~Fits 1 \& 2 used all 18 sources in Table~\ref{table:parallaxes} and
have high \chisq\ values, owing to two outliers: NGC~7538 and G~23.6$-$0.1.  
Fit~3 excludes the two outliers and provides our basic result, under the assumption of a 
flat rotation curve.  Fits 4 -- 7 explore the effects of non-flat rotation curves.
``DF'' is the degrees of freedom for the fit (\ie number of data equations minus number 
of parameters).  (\Usbar,\Vsbar,\Wsbar) are average peculiar motions common to all sources 
(see Table~\ref{table:definitions} and Figure~\ref{fig:definitions}), assuming the Hipparcos 
solar motion of \citet{Dehnen:98} (see discussion in \S\ref{sect:3-D}).
All \To/\Ro\ estimates were obtained by holding $\Ro=8.50$~kpc and solving for \To.  
``Clemens-10'' and ``Clemens-8.5'' 
refer to the \cite{Clemens:85} rotation curves for (\Ro[kpc],\To[\kms]) = (10,250) 
and (8.5,220), respectively; ``Brand-Blitz'' refers to the \citet{Brand:93} rotation
curve.  Both the Clemens and Brand-Blitz rotation curves were scaled to the fitted
values of \Ro\ and \To.   
               }
\label{table:fits}
\end{deluxetable}

\subsection {Rotation Curves} \label{sect:rotation_curves}

We have until now assumed that the rotation curve of the Galaxy is flat
(\ie $\Theta(R)=\To$).  In order to investigate deviations from a flat rotation curve, 
we used the basic sample of 16 sources and added the parameter \Tdot\ to the model.  
A least-squares fit yielded $\Tdot=2.3\pm0.9$~\kmsperkpc, with an improved \chisq\ 
compared to the flat rotation curve fit (see Fit~4 in Table~\ref{table:fits}), 
but with an increased correlation coefficient between \Ro\ and \To\ of 0.90.  
We tested how sensitive $\Tdot$ was to the two outer Galaxy sources by dropping
S~269 and WB~89-437 from the sample and re-fitting.  This yielded 
$\Tdot=1.9\pm1.2$~\kmsperkpc\ and indicated that these sources do not provide all 
the leverage for a rising rotation curve.  Thus, we find a nearly flat rotation 
curve between Galactocentric radii of about 4 to 13~kpc, with some evidence for a 
slight rise with distance from the Galactic center.  This supports similar 
conclusions reached in a number of papers 
\citep{Fich:89,Brand:93,Honma:97,Maciel:05}.  
For example \citet{Fich:89} find that the rotation curve is nearly
flat for $\To=220$~\kms\ and that it rises gradually for $\To=250$~\kms.

We also tested more complex rotation curves by replacing the simple
linear form just discussed with the rotation curves of \citet{Clemens:85}
and \citet{Brand:93}.
Clemens supplied two curves: one assuming the old IAU constants of
$\Ro=10$~kpc and $\To=250$~\kms\ and the other assuming the new constants
of $\Ro=8.5$~kpc and $\To=220$~\kms.  These models have slightly different
shapes, with the old model generally having rotational speeds that rise faster 
with radius than the new model.   For either model, we fitted for
different values of \Ro\ (which we used to scale model radii) and \To\ 
(which we used to scale rotation speeds).  The fit using the old model,
listed as Fit~5 in Table~\ref{table:fits}, gave 
$\Ro=8.7\pm0.4$~kpc and $\To=272\pm15$~\kms, with an improved $\chisq=52.9$ for 59
degrees of freedom compared to our solution for a flat rotation curve.
The improvement is partly from a better match to the two sources in the
Outer arm (S~269 and WB~89-437).  Using the new rotation model, gave
$\Ro=7.9\pm0.3$~kpc and $\To=230\pm12$~\kms, with a considerably worse $\chisq=71.2$
(see Fit~6 in Table~\ref{table:fits}).  
Using the \citet{Brand:93} rotation curve, also scaled by the fitted values
of \Ro\ and \To, we obtain Fit~7 in Table~\ref{table:fits}, with values of
$\Ro=8.8\pm0.4$~kpc and $\To=275\pm15$~\kms\ and
a $\chisq=59.0$ for 59 degrees of freedom, intermediate between the
$\chisq$ values for the two Clemens models.

Clearly there is some sensitivity of the best fit \Ro\ value to the models,
and we adopt a systematic uncertainty in \Ro\ of $\pm0.5$~kpc. 
Note that, as discussed in \S\ref{sect:gal_model}, the ratio $\To/\Ro$ has much
less modeling sensitivity.
With the current parallax and proper motion data, we cannot conclusively distinguish 
among the rotation curves presented.   However, with the many more parallaxes
and proper motions expected in the next few years from the VLBA and VERA telescopes, 
we should be able to make considerable progress in refining the rotation curve 
of the Milky Way.

\section{Kinematic Distances} \label{sect:k_dist}

\begin{figure}[htp]
\epsscale{1.0} 
\plotone{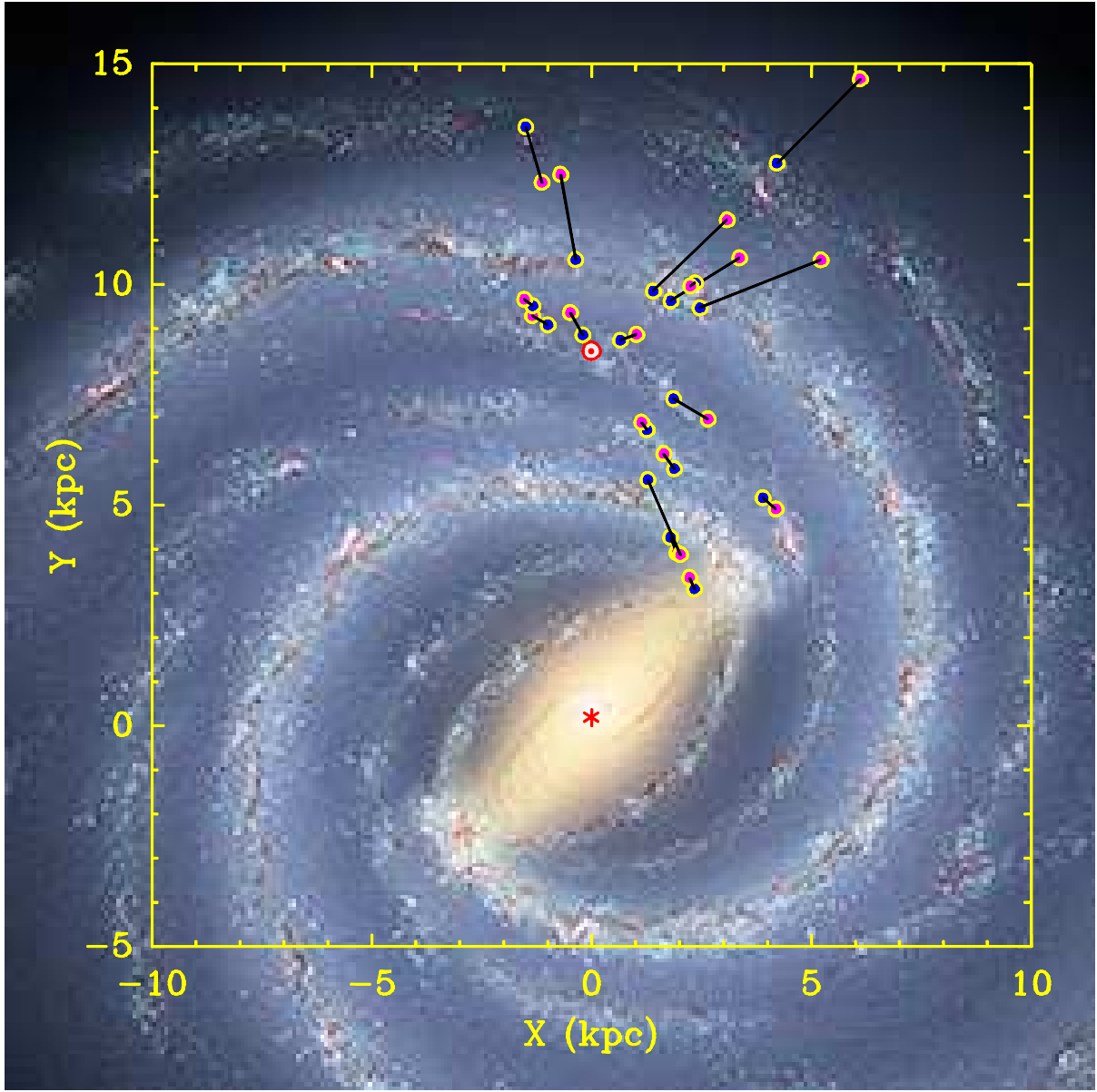} 
\caption{\small
The locations of the star forming regions determined by trigonometric 
parallax ({\it dark blue circles)} and by kinematic distances 
({\it light magenta circles}),
assuming IAU recommended values of $\Ro=8.5$~kpc and $\To=220$~\kms\ 
and the Standard Solar Motion to define the LSR.  
        }
\label{fig:k_distances}
\end{figure}

Figure~\ref{fig:k_distances} compares the locations of the star forming
regions determined by trigonometric parallax and by kinematic distances.
The kinematic distances were computed for the IAU standards
$\Ro=8.5$~kpc and $\To=220$~\kms\ and the standard definition of LSR.
For 13 of 18 regions (11 of 16 in the basic sample), the kinematic distance exceeds 
the true source distance; in 3 cases the discrepancy is over a factor of two.
The kinematic distances for (these) star forming regions tend to
over-estimate the source distances.

As shown above, HMSFRs on average orbit the Galaxy $\approx15$~\kms\ 
slower than the circular rotation speed.  Taking this into account, a prescription for 
a ``revised'' kinematic distance for a high-mass star forming region could be as follows:
\begin{itemize} 
\item [1)] add back the (old) Standard Solar Motion corrections to the LSR 
   velocities, returning them to the heliocentric frame;
\item [2)] apply ``best values'' for the solar motion to calculate a revised 
   ``LSR'' velocity, \vlsrr;
\item [3)] subtract $-15$~\kms\ from the velocity component in the direction of
   Galactic rotation;
\item [4)] calculate a kinematic distance using values for the 
   fundamental parameters of the Milky Way, \eg  $\Ro=8.4$~kpc and 
   $\To=254$~\kms, that are consistent with astrometric measurements; and
\item [5)] when determining the {\it uncertainty} in the kinematic distance, 
   include a systematic contribution allowing for the possibility of a
   $7$~\kms\ uncertainty in \vlsrr.
\end{itemize}

Table~\ref{table:k_distances} gives parallax distances, standard (old)
kinematic distances, and revised kinematic distances and uncertainties
(using the above prescription) for all 18 HMSFRs listed in Table~\ref{table:parallaxes}.
(We provide the FORTRAN source code used to calculate revised kinematic distances in 
the on-line material.)
Note that our prescription for the uncertainty in kinematic distances
performs reasonably well for our basic sample (excluding the two sources G~23.6$-$0.1 and
NGC~7538 which we earlier noted as outliers).  The mean difference between the
parallax and kinematic distances is near zero and the differences divided by their 
uncertainties average to near unity.  Now only half (8 of 16) of the sources 
in the basic sample have kinematic distances that exceed the true source distance.
There are no cases for which the discrepancy is a factor of two, 
and the estimated uncertainties reasonably account for differences between the
parallax and kinematic distances.

\begin{deluxetable}{lrrrrrr}
\tablecolumns{7} \tablewidth{0pc}
\tablecaption{Parallaxes vs. Kinematic Distances}
\tablehead {
  \colhead{Source} & \colhead{$\ell$} & \colhead{$b$} & \colhead{\vlsr} &
  \colhead{$D_\pi$} & \colhead{$D_k^{Std}$} & \colhead{$D_k^{Rev}$} 
\\
  \colhead{}      & \colhead{(deg)} & \colhead{(deg)} & \colhead{\kms} &
  \colhead{(kpc)} & \colhead{(kpc)} & \colhead{(kpc)} 
           }
\startdata
G~23.0$-$0.4...& 23.01 &$-0.41$ &$+81$ &4.59  &4.97  &$4.72^{+0.3}_{-0.3}$ \\
G~23.4$-$0.2...& 23.44 &$-0.18$ &$+97$ &5.88  &5.60  &$5.29^{+0.3}_{-0.3}$ \\
G~23.6$-$0.1...& 23.66 &$-0.13$ &$+83$ &3.19  &5.04  &$4.77^{+0.3}_{-0.3}$ \\
G~35.2$-$0.7...& 35.20 &$-0.74$ &$+28$ &2.19  &2.00  &$1.99^{+0.4}_{-0.4}$ \\
G~35.2$-$1.7...& 35.20 &$-1.74$ &$+42$ &3.27  &2.85  &$2.76^{+0.4}_{-0.4}$ \\
W~51~IRS~2..   & 49.49 &$-0.37$ &$+56$ &5.13  &5.52  &$5.46^{+1.6}_{-1.6}$ \\
G~59.7$+$0.1...& 59.78 &$+0.06$ &$+27$ &2.16  &3.07  &$3.45^{+1.2}_{-1.2}$ \\
Cep~A...........&109.87&$+2.11$ &$-10$ &0.70  &1.09  &$0.55^{+0.7}_{-0.6}$ \\
NGC~7538....   &111.54 &$+0.78$ &$-57$ &2.65  &5.61  &$4.64^{+0.7}_{-0.6}$ \\
IRAS~00420..   &122.02 &$-7.07$ &$-44$ &2.13  &3.97  &$3.18^{+0.6}_{-0.6}$ \\
NGC~281......  &123.07 &$-6.27$ &$-31$ &2.82  &2.69  &$2.08^{+0.6}_{-0.6}$ \\
W3(OH).......  &133.95 &$+1.06$ &$-45$ &1.95  &4.28  &$3.42^{+0.7}_{-0.7}$ \\
WB~89-437...   &135.28 &$+2.80$ &$-72$ &5.99  &8.68  &$6.89^{+1.2}_{-1.0}$ \\
S~252............&188.95&$+0.89$&$+11$ &2.10  &4.06  &$3.33^{+4.2}_{-2.4}$ \\
S~269............&196.45&$-1.68$&$+20$ &5.29  &4.13  &$3.35^{+2.0}_{-1.5}$ \\
Orion............&209.01 &$-19.38$&$+10$&0.41 &0.99  &$0.71^{+0.7}_{-0.6}$ \\
G~232.6$+$1.0..&232.62 &$+1.00$ &$+23$ &1.68  &1.92  &$1.44^{+0.6}_{-0.5}$ \\
VY~CMa.......  &239.35 &$-5.06$ &$+18$ &1.14  &1.56  &$1.10^{+0.6}_{-0.6}$ \\
\enddata
\tablecomments {\footnotesize
$D_\pi$ is the measured parallax converted to distance;
$D_k^{Std}$ is the kinematic distance based on standard LSR velocities; 
$D_k^{Rev}$ and $\sigma(D_k)$ is the revised kinematic distance and
its uncertainty, calculated for $\Ro=8.4$~kpc, $\To=254$~\kms\ and $\Usbar=-15$~\kms, 
following the prescription outlined in \S\ref{sect:k_dist}.  
All kinematic distances assume a flat rotation curve.
               }
\label{table:k_distances}
\end{deluxetable}

While the prescription outlined above results in some improvement in kinematic 
distances compared to the standard approach, the improvement is not as great as 
one might at first expect.  This occurs because the definition
of the LSR uses the Standard Solar Motion.  While the Standard Solar Motion
differs only slightly from the Hipparcos solar motion for components
toward the Galactic center (\U) and the north Galactic pole (\W),
there is a large discrepancy for the component in the direction of Galactic 
rotation.  The Standard value is $\Vo=15.3$~\kms, whereas the Hipparcos value is
$\VH=5.25$~\kms.  The $+10$~\kms\ ``error'' in \Vo\ partially
compensates for the $15$~\kms\ slower Galactic orbits of HMSFRs
shown in \S\ref{sect:3-D}.  (Note that a positive change in the solar motion component 
\V\ results in a negative change in a source peculiar motion component \Vsbar).
Even with the improved prescription for kinematic distances, 
one cannot really hope to discern spiral structure using kinematic distances.

\section{Galactic Coordinates} \label{sect:gal_coordinates}
 
There is excellent agreement between the two independent VLBI measurements of $\To/\Ro$:
the measurement based on parallaxes and proper motions of HMSFRs (this paper) and 
based on the proper motion of Sgr~A* \citep{Reid:04}.  This gives us confidence that 
1) we can well model the Galaxy with parallax and proper motions of HMSFRs and 
2) Sgr~A* is indeed a supermassive black hole at the dynamical center of the Milky Way.  
These findings offer an independent definition of the Galactic plane and Galactic
coordinates.  Currently, the IAU definition of the Galactic plane is based primarily 
on the thin distribution of neutral hydrogen 21~cm emission \citep{Blaauw:60}.  
The Sun is defined to be precisely in the plane and the origin 
of longitude was set by the centroid of the radio emission of the large, 
complex source Sgr~A.   The Sun is now known to be $\approx20$~pc north of the plane 
(see \citet{Reed:06} and references therein) and the supermassive black hole, 
Sgr~A*, is offset by a few arcmin from the IAU defined center.  

In the future, one could consider redefining Galactic coordinates based,
in part, on the proper motion of Sgr~A*, which, after correction for the well-determined 
solar motion component perpendicular to the Galactic plane, gives the orbital plane of the
Local Standard of Rest.  The zero of longitude would be best defined by the 
position of Sgr~A*.   This would place our supermassive black hole at the origin 
of Galactic coordinates, and one could rotate the reference frame to remove the Sun from
its special location precisely in the Galactic plane.  

\section{Discussion}   \label{sect:discussion}

Very Long Baseline Interferometry now routinely yields parallax measurements 
with accuracies of $\sim10$~\uas, corresponding to 10\% uncertainty at a distance 
of 10~kpc, and proper motions that are usually accurate to $\sim 10$~\uasy\ or 
better than than $\sim1$~\kms\ at similar distances.  
Target sources include molecular masers associated 
with star formation and red giant stars, as well as non-thermal continuum emission 
associated with young T~Tau stars and cool dwarfs.   
Combining the first results of parallaxes for high-mass star forming
regions from the VLBA and the Japanese VERA project has allowed us to
begin to investigate the spiral structure and kinematics of the Galaxy.

We have accurately located three of the spiral arms of the Milky Way and directly 
measured a pitch angle of 16\deg\ for a portion of the Perseus spiral arm.  
This pitch angle is similar to those of spiral arms in other galaxies of 
type Sb to Sc \citep{Kennicutt:81}.  Two armed spirals can account for most of the
known large H~II regions only if the arms wrap twice around the Galaxy;  
this requires pitch angles of $\approx8\deg$.  With a pitch angles greater than 
$\approx12\deg$, the Galaxy needs to have four arms in order to account for the 
approximate locations of H~II regions \citep{Georgelin:76,Taylor:93}.   
There has been considerable discussion in the literature concerning the number of
spiral arms in the Galaxy \citep{Simonson:76,Bash:81,Vallee:95,Drimmel:00,Drimmel:01,
Benjamin:05,Nakanishi:06,Steiman:08}, with Spitzer GLIMPSE survey results suggesting 
only two arms can be traced in the redder, older population of stars \citep{Benjamin:08}.
Perhaps the VLBI and infrared survey results can be reconciled if the Milky Way 
exhibits a hybrid structure, consisting of two dominant spiral arms, populated by 
both young and old stars and with pitch angles near $16\deg$, and two weaker arms 
traced only by young stars. 

Our finding that HMSFRs on average orbit the Galaxy $\approx15$~\kms\ slower than 
expected for circular orbits has implications for star formation and spiral density 
wave theory.  The plot of the {\it apparent} solar motion in the direction of Galactic
rotation (\V) versus stellar dispersion (Figure~\ref{fig:hipparcos}) can be 
interpreted as a time sequence, with stellar age increasing with dispersion.  
The 15~\kms\ slower orbital speed of HMSFRs displays as a positive departure of the 
apparent Solar Motion with respect to the asymmetric drift (the fitted trend to the 
Hipparcos data shown in Figure~\ref{fig:hipparcos}), since the Sun {\it appears} to 
orbit faster when measured against such stars.  One explanation for this finding is 
that HMSFRs are born in elliptical Galactic orbits, near apocenter, with orbital 
eccentricity of about 0.06.  As young stars continue to orbit the Galaxy, their 
orbits become more circularized, as evidenced by the lesser departure of the 
youngest Hipparcos bin (mostly late B-type stars) from the asymmetric drift line 
compared to the HMSFRs.  The gradual transfer of angular momentum from gas to stars 
in the Galaxy proposed by \citet{Chakrabarti:09} may explain this.  At a stellar 
dispersion of $\approx300$~(\kms)$^2$, which corresponds to A2- to A5-type stars 
with colors $B-V=0.1$ and characteristic main-sequence lifetimes of $\sim1$~Gy, the 
stars join the asymmetric drift.  As stars continue to age, their orbits are 
progressively ``randomized'' and  they (again) become part of a slower orbiting 
population, which appears as a larger apparent \V.

Parallaxes measurements alone generally cannot yield \Ro\ (except for a parallax of 
\SgrA).  However, since galaxies rotate in a fairly smooth fashion, 
a kinematic model can be directly compared with distance and relative motion 
measurements in order to estimate \Ro\ and \To.  
In this paper, we have demonstrated that parallax and proper motion measurements 
for HMSFRs across large portions of the Galaxy can separate estimates
of \Ro\ and \To, although because of the somewhat restricted coverage of the Galaxy
currently available, we have a significant correlation between these
parameters.   Our best estimate of \Ro\ is $8.4\pm0.36\pm0.5$~kpc, where the second 
uncertainty is systematic and comes from our lack of detailed knowledge
of the rotation curve of the Galaxy.   This estimate is consistent with the ``best'' 
\Ro\ of $8.0\pm0.5$~kpc from a combination of many methods reviewed by \citet{Reid:93}.  
Also, recent {\it direct} estimates of \Ro\ from radial velocities and elliptical 
paths of stars that orbit \SgrA\ have converged on values of $8.4\pm0.4$~kpc
\citep{Ghez:08} and $8.33\pm0.35$ \citep{Gillessen:09}.  (These estimates assume
that \SgrA\ is nearly motionless at the Galactic center.  Relaxing this assumption
decreases the estimates to about 8.0~kpc.)  Of course, many other less direct 
estimates of \Ro\ can be found in the literature and span a much greater range.

The characteristic rotation speed of the Galaxy (\To) is a crucial parameter not only
for Galactic dynamics and kinematic distance determinations, but also for estimating 
the total mass in dark matter and the history and fate of the Local Group of galaxies 
\citep{Loeb:05,Shattow:08}.
Estimates of the rotation speed of the Galaxy from the recent literature span
a very large range between 184~\kms\ \citep{Olling:98} and 272~\kms\ \citep{Mendez:99}.
Most estimates of \To\ are based on analyses of the shear and vorticity of large samples
of stars in the (extended) solar neighborhood and thus really measure Oort's A and B parameters.
Quoted values of \To\ then come by {\it assuming} a value for \Ro\ and using the relation
$\To = \Ro (A-B)$.  Our result that $\To=254\pm16$~\kms\ was obtained by fitting for
both \Ro\ and \To\ using full 3-dimensional locations and motions of sources well beyond the 
extended solar neighborhood and, thus, does not assume a value for \Ro.  However, as 
discussed in \S\ref{sect:gal_model}, with the present distribution of sources there is 
considerable correlation between \Ro\ and \To\ parameters, which is reflected in the 
$\pm16$~\kms\ formal uncertainty for \To.

Our estimate of the ratio $\To/\Ro$ of $30.3\pm0.9$~\kmsperkpc\ is determined more
accurately than either parameter individually and is nearly independent of the 
value of \Ro\ over the range of likely values between about 8.0 and 8.5~kpc.  
This value differs considerably from that determined from the IAU values of 
$\To/\Ro=220~{\rm km~s}^{-1}/8.5~{\rm kpc}=25.9$~\kmsperkpc\ and differs marginally from 
the \citet{Feast:97} analysis of Hipparcos Cepheids of $27.19\pm0.87$~\kmsperkpc.
Recent studies, using samples of OB-type stars within 3~kpc of the Sun (excluding Gould's 
Belt stars and based on Hipparcos data augmented with photometrically determined distances), 
arrive at $A-B$ between 30~\kmsperkpc\ \citep{Uemura:00} and 32~\kmsperkpc\ 
\citep{Miyamoto:98,Elias:06}, with uncertainties of about $\pm1.5$~\kms. 

Our value for $\To/\Ro$ is in excellent agreement with that determined directly from 
the {\it apparent} proper motion of Sgr~A* (the supermassive black hole at the Galactic center)
of $6.379\pm0.024$~\masy\ \citep{Reid:04}.
One expects a supermassive black hole to be stationary at the dynamical center of the 
Galaxy to better than $\sim1$~\kms\ \citep{Chatterjee:02,Dorband:03,Reid:04}. 
Hence, Sgr~A*'s {\it apparent} motion should be dominated by the effects of the Galactic 
orbit of the Sun.  After correcting for the Solar Motion of 5.25~\kms\ in the
direction of Galactic rotation \citep{Dehnen:98}, Sgr~A*'s apparent motion yields 
a global estimate of $\To/\Ro=29.45\pm0.15$~\kmsperkpc.  Thus, there is excellent 
agreement between this and our global and direct method for measuring $\To/\Ro$.  

Coupling the \SgrA\ motion result of $\To/\Ro=29.45\pm0.15$~\kmsperkpc\ with 
estimates of \Ro\ from stellar orbits in the Galactic center \citep{Ghez:08,Gillessen:09}
of $8.4\pm0.4$~kpc yields $\To=247\pm12$~\kms.  This result is 
also a direct and global measurement of $\To$ and is independent of our result
from parallaxes and proper motions of star forming regions.  Combining the
Galactic center and star forming region estimates gives $\To=250\pm10$~\kms.  

It seems clear that \To\ is near the upper end of the range of estimates in the 
literature.  We note that both the Galactic center stellar orbit and the star forming 
region parallax results assume the Hipparcos solar motion of $5.25$~\kms\ in the 
direction of Galactic rotation.  Only if the interpretation of the asymmetric drift 
is incorrect or if the entire solar neighborhood orbits the Galactic center $\sim30$~\kms\ 
slower than the Galaxy spins could $\To$ be equal to the IAU recommended value of 
220~\kms.

\begin{figure}[htp]
\epsscale{1.0} 
\plotone{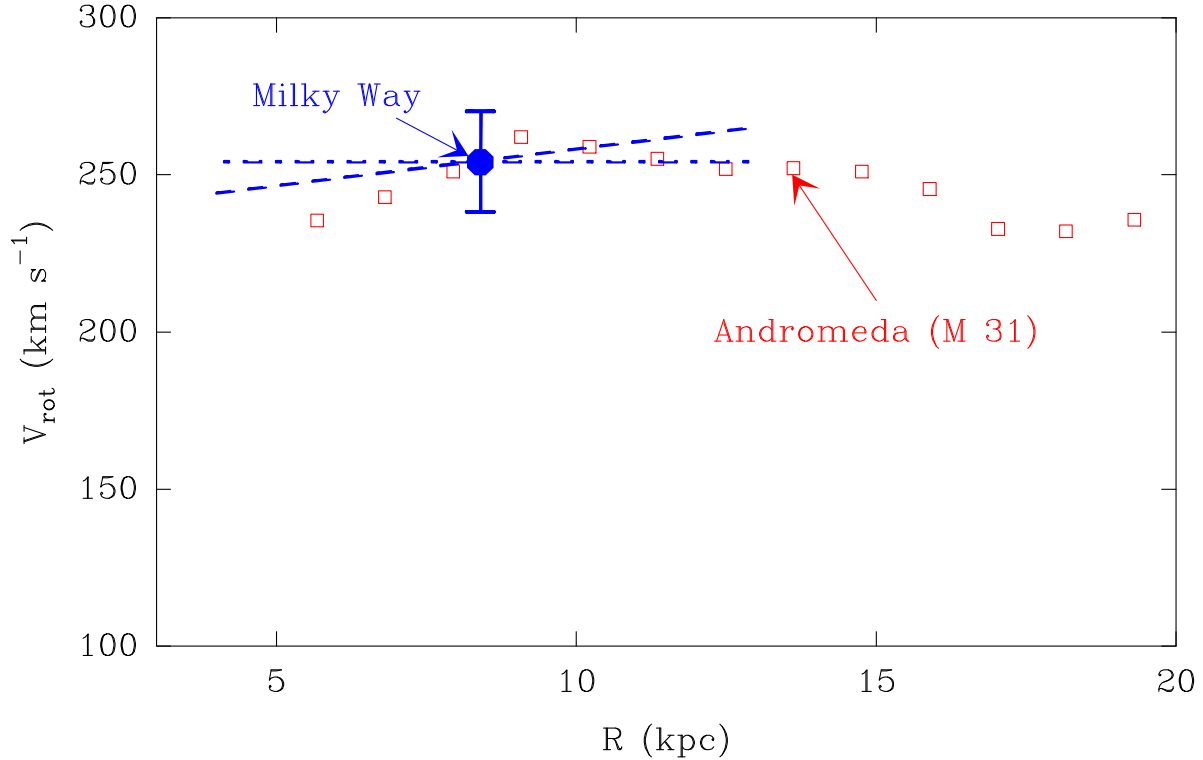} 
\caption{\small
Rotation speed versus radius for the Andromeda galaxy and the Milky Way.
The {\it red squares} are based on HI observations of Andromeda tabulated by \citet{Carignan:06}.
The {\it blue filled circle} is our best estimate of $\To=254\pm16$~\kms\ at $\Ro=8.4$~kpc
for the Milky Way, derived from the parallax and proper motions of high mass star 
forming regions.  The {\it blue dot-dashed line} is for a flat rotation curve, and
the {\it blue dashed line} corresponds to a slightly rising rotation curve of
2.3~\kmsperkpc (see \S\ref{sect:rotation_curves}).  These lines are plotted
over the range of Galactocentric radii sampled by the parallax and proper motion results.  
Note that these two galaxies have nearly identical rotation speeds over this range.
        }
\label{fig:andromeda}
\end{figure}

We have determined the rotation speed of the Milky Way at the radius of the Sun 
to be $\approx250$~\kms\ and the rotation curve to be nearly flat or slightly rising 
with distance from the Galactic center.  These values are nearly identical to those of 
the Andromeda galaxy (M31) as shown in Figure~\ref{fig:andromeda}.  
The rotation curve of Andromeda, determined from HI emission by 
\citet{Carignan:06} based on interferometric observations of \citet{Unwin:83}, 
indicates a speed of 251~\kms\ at a radius of 8~kpc, a slightly 
rising curve out to about 15~kpc, and a slow dropoff to about 225~\kms\ beyond 20~kpc.   
The most straightforward interpretation of the similarities of the rotation curves for 
the Milky Way and Andromeda is that these two galaxies are nearly equal in size and mass.  
  
Finally, we note that \citet{Reid:04} placed a strong upper limit of $-0.4\pm0.9$~\kms\ 
for the component of peculiar motion of \SgrA\ {\it perpendicular} to the plane of the Galaxy.  
However, the determination of the component in the direction of Galactic rotation was 
considerably less accurate: $18\pm7$~\kms, as one must remove the uncertain effects of
the solar orbit.  Reid \& Brunthaler did this by removing $27.19\pm0.87$~\kmsperkpc, based on 
Hipparcos measurements of Oort's constants ($A-B$) by \citet{Feast:97}, from the observed
motion of Sgr~A* in the Galactic plane of 29.45~\kmsperkpc.   This method assumes that 
$\To/\Ro=A-B$ and that estimates of the shear and vorticity of nearby stars 
from Hipparcos data indicate the large-scale differential rotation of the Galaxy and 
are not subject to local irregularities in the solar neighborhood.
Since we now have a direct, global estimate of  $\To/\Ro=30.3\pm0.9$~\kmsperkpc, we find the 
peculiar motion of \SgrA\ in the direction of Galactic rotation to be $-7.2\pm8.5$~\kms,
with little sensitivity to \Ro\ (adopted to be 8.5~kpc here).  This adds additional 
strong evidence that \SgrA\ is a supermassive black hole, which is nearly stationary 
at the dynamical center of the Galaxy.

\vskip 1.0truein 
XWZ, BZ, and YX were supported by the Chinese National Science Foundation, 
through grants NSF 10673024, 10733030, 10703010 and 10621303, and by the NBPRC 
(973 Program) under grant 2007CB815403.

\section{Appendix}

     Since we have measured the position, distance, LSR velocity and proper motion 
of each source, we know its full 3-dimensional location in the Galaxy and full 
space motion relative to the Sun.  Given a model of Galactic rotation, we can then 
calculate the non-circular (peculiar) velocity of each source.  While this calculation 
is conceptually simple, in practice, there are some subtleties and sign convention 
issues that can lead to errors, and so here we present the necessary formulae (and
FORTRAN source code in the on-line material).

\begin{figure}[htp]
\epsscale{0.8} 
\plotone{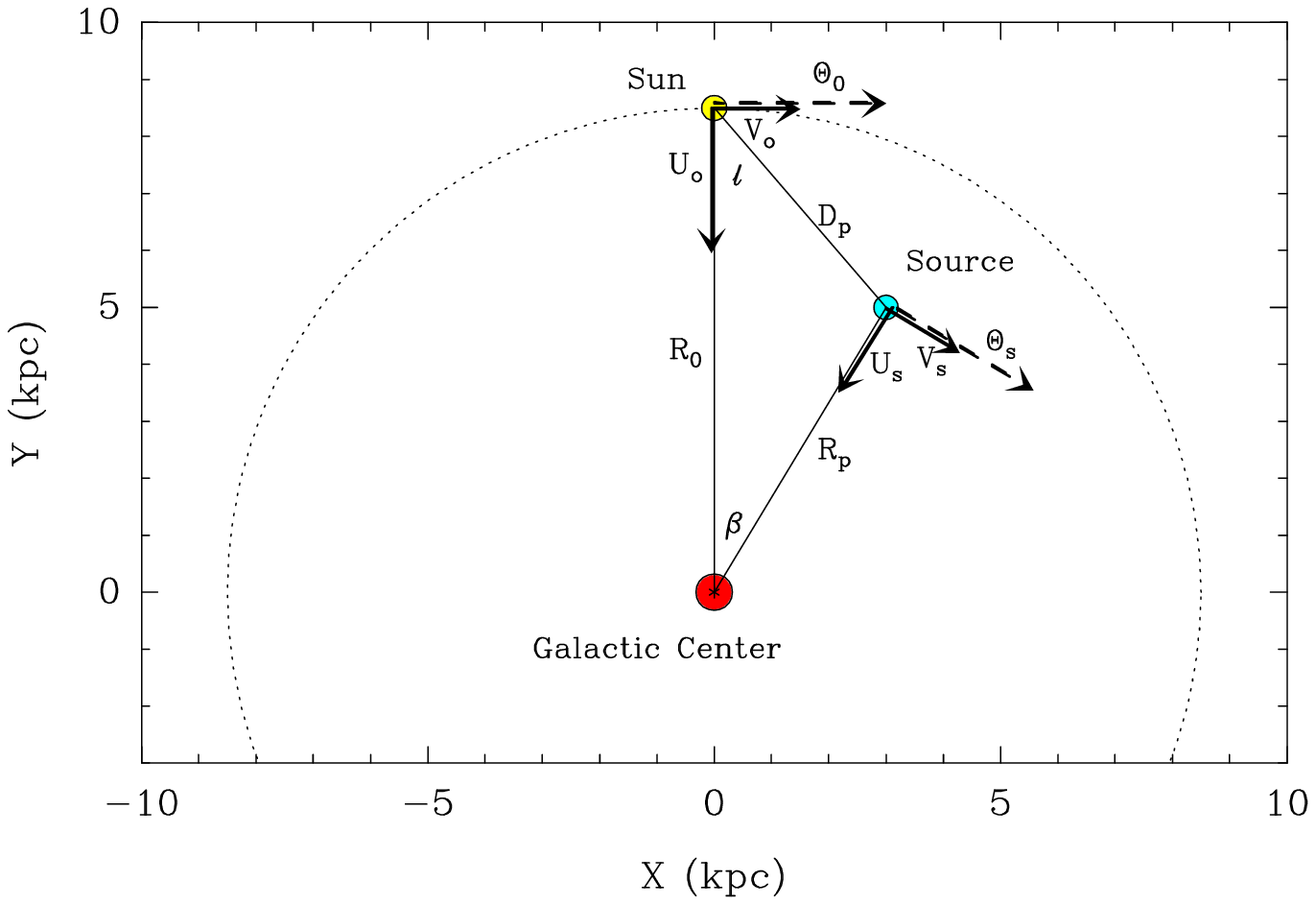} 
\caption{\small
Schematic depiction of source and Galactic parameters.
        }
\label{fig:definitions}
\end{figure}

The required Galactic and Solar Motion parameters are given in
Table~\ref{table:parameters}, and those associated with the
source are defined in Table~\ref{table:definitions}.  A schematic 
depiction of these parameters is given in Figure~\ref{fig:definitions}.
We assume that the Sun is in the Galactic plane 
and calculate a source's peculiar motion (\ie with respect to 
a circular Galactic orbit) as follows:

\begin{deluxetable}{lrl}
\tablecolumns{3} \tablewidth{0pc} 
\tablecaption{Galactic and Solar Parameters and Nominal Values}
\tablehead {
  \colhead{Parameter} & \colhead{Value} &   \colhead{Definition} 
            }
\startdata
\Ro ......& 8.5~kpc & Distance to the GC (IAU value)\\
\To ......& 220~\kms& Rotation speed of LSR (IAU value)\\
\Ts ......& 220~\kms& Rotation speed of Galaxy at source\\
\\
\Uo ......& 10.3~\kms  & Standard Solar Motion toward GC\\
\Vo ......& 15.3~\kms  & Standard Solar Motion toward $\ell=90\deg$ \\
\Wo ......&  7.7~\kms  & Standard Solar Motion toward NGP\\
\\
\UH ......& 10.0~\kms  & Hipparcos Solar Motion toward GC\\
\VH ......&  5.2~\kms  & Hipparcos Solar Motion toward $\ell=90\deg$ \\
\WH ......&  7.2~\kms  & Hipparcos Solar Motion toward NGP\\
\enddata
\tablecomments {\footnotesize
 GC: the Galactic Center; LSR: Local Standard of Rest;
 NGP: North Galactic Pole.  The Standard Solar Motion must be used to
 convert from \vlsr\ to \vhelio, since (hopefully) all observatories
 have used this definition.  The values given above
 come from an assumed Solar Motion of 20~\kms\ toward R.A.(1900)=18$^h$ 
 and Dec.(1900)-30$^\circ$ precessed to J2000.0.
 Hipparcos Solar Motion values are from \citet{Dehnen:98}.
               } 
\label{table:parameters}
\end{deluxetable}

\begin{deluxetable}{ll}
\tablecolumns{2} \tablewidth{0pc} 
\tablecaption{Source Parameter Definitions}
\tablehead {
  \colhead{Parameter} & \colhead{Definition} 
            }
\startdata
$\ell$ ..........& Galactic longitude \\
$D$ ........ & Distance from Sun ($1/\pars$)\\
$D_p$ ....... & Distance from Sun projected in plane\\
\Rp\ ....... & Distance from GC projected in plane \\
\vlsr\ ....& LSR radial velocity  \\
\vhelio\ ...& Heliocentric radial velocity  \\
\ura\ ....... & Proper motion in R.A. ($\ura=\mu_x/\cos\delta$) \\
\udec\ .......& Proper motion in Decl. ($\udec=\mu_y$) \\
$\beta$ ......... & Angle: Sun--GC--source   \\
\Us\ ....... & Peculiar motion locally toward GC  \\
\Vs\ ....... & Peculiar motion locally in direction of Galactic rotation\\
\Ws\ ...... & Peculiar motion toward NGP \\
\enddata
\tablecomments {\footnotesize
 GC: the Galactic Center; LSR: Local Standard of Rest;
 NGP: North Galactic Pole.  
               } 
\label{table:definitions}
\end{deluxetable}

We convert \vlsr\ to a heliocentric frame, $\vhelio$, by adding back the
component of the Standard Solar Motion in the line-of-sight direction
that had been removed from the observed Doppler shift to calculate \vlsr.
Note that one needs to use the (old) Standard Solar Motion, which 
defines the LSR frame, and {\it not} the best values available today.
Generally, observatories have adopted a value of 20~\kms\ toward
$\alpha(1900)=18^{\rm h}, \delta(1900)=+30^{\rm d}$ for the 
Standard Solar Motion.  Precessing these coordinates to the 
epoch of observation ($\approx2006$) and converting to Galactic 
Cartesian coordinates yields the (\Uo,\Vo,\Wo) values listed
in Table~\ref{table:parameters}.  Then,
$$\vhelio = \vlsr - ( \Uo \cos{\ell} + \Vo \sin{\ell} )\cos{b}
     - \Wo \sin{b}~~.$$

Rotate the motion vector from the equatorial heliocentric frame 
(\ura,\udec,\vhelio) to a Galactic heliocentric frame (\ul,\ub,\vhelio).  
This is a rotation about a radial axis and is defined by the IAU in B1950 
coordinates \citep{Blaauw:60}.
For coordinates in J2000, \citet{Reid:04} give the right ascension and
declination of the NGP as $\alpha_P=12^h51^m26\s2817$ and 
$\delta_P=27\deg07'42\as013$, respectively, and the zero of longitude
is the great semicircle originating at the NGP at the position angle
$\theta = 122\d932$.
Galactic latitude can be obtained from
$$\sin{b} = \sin{\delta} \cos{(90\deg-\delta_P)} -
        \cos{\delta} \sin{(\alpha-\alpha_P-6^h)} \sin{(90\deg-\delta_P)}~~.$$
\noindent
A useful angle $\phi$ can be determined (between 0\deg and 360\deg) from
$$\sin{\phi} =\biggl(\cos{\delta} \sin{(\alpha-\alpha_P-6^h}) \cos{(90\deg-\delta_P)}
        + \sin{\delta} \sin{(90\deg-\delta_P)} \biggr) / \cos{b} $$
\noindent
and
$$\cos{\phi} = \cos{\delta} \cos{(\alpha-\alpha_P-6^h)} / \cos{b}~~, $$
\noindent
and then Galactic longitude follows from
$$\ell = \phi + (\theta - 90\deg)~~.$$

Proper motion in Galactic coordinates ($\ul,\ub$) can be easily calculated 
from the motion in equatorial coordinates ($\ura,\udec$) by differencing 
($\ell,b$) values for coordinates determined, say, one year apart.
This usually requires 64-bit precision in the calculations.
Note that \ul\ will naturally be defined positive in the
direction of increasing Galactic longitude, which is {\it counter} to
Galactic rotation.

Convert the proper motions to linear speeds (by multiplying by distance) via
$$v_\ell = D \ul\cos{b}~~~{\rm and}~~~v_b = D \ub ~~,$$
where $\ul\cos{b}$ is the actual motion tangent to the sky in the direction
of Galactic longitude.

We now convert from spherical to Cartesian Galactic coordinates at the
location of the Sun.
$$U_1 = ( \vhelio \cos{b} - v_b \sin{b} )\cos{\ell} - v_\ell \sin{\ell}~~,$$
$$V_1 = ( \vhelio \cos{b} - v_b \sin{b} )\sin{\ell} + v_\ell \cos{\ell}~~,$$
$$W_1 = v_b \cos{b} + \vhelio \sin{b}~~.$$

Next add the full orbital motion of the Sun, using the best values of
the solar motion and the circular rotation of the Galaxy at the position
of the Sun (\UH,\VH+\To,\WH):
$$U_2 = U_1 + \UH ~~,~~~V_2 = V_1 + \VH + \To~~,~~~W_2 = W_1 + \WH~~.$$

The Galactocentric distance to the source projected onto the Galactic plane is given by
$$ \Rp^2 = \Ro^2 + D_p^2 - 2\Ro D_p \cos{\ell}~~, $$
where $D_p=D \cos{b}$.
The angle $\beta$ between the Sun and the source as viewed from the Galactic 
center can be determined in all cases (\ie from 0\deg\ to 360\deg) from
$$ \sin{\beta} = {D_p\over \Rp} \sin{\ell}~~~~~~{\rm and}~~~~~~
    \cos{\beta} = { {\Ro - D_p\cos{\ell}} \over \Rp}~, $$
Rotate the vector $(U_2,V_2,W_2)$ through the angle $\beta$ in the 
plane of the Galaxy and remove circular Galactic rotation at the location
of the source to yield $(\Us,\Vs,\Ws)$ :
$$\Us = U_2 \cos{\beta} - V_2 \sin{\beta}~~,$$
$$\Vs = V_2 \cos{\beta} + U_2 \sin{\beta}~- \Ts~~,$$
$$\Ws = W_2~~.$$

The vector (\Us,\Vs,\Ws) gives the non-circular (peculiar) motion
of the source in a Cartesian Galactocentric frame, where \Us\ is radially inward
toward the Galactic center (as viewed by the source), \Vs\ is in the local
direction of Galactic rotation and \Ws\ is toward the north Galactic pole.

\end{document}